\documentclass[aps,prx,amsmath,twocolumn,amssymb,notitlepage,superscriptaddress]{revtex4-1}	
\usepackage[english]{babel}
\usepackage{graphicx}
\usepackage{amsmath}
\usepackage{amssymb}
\usepackage{epstopdf}
\usepackage{amsfonts}
\usepackage{bm}
\usepackage{times}
\usepackage{mathtools}
\usepackage{hyperref}

\usepackage{color}

\newcommand{\astcycl}{\mathrlap{\kern0.085em{\circlearrowright}}\ast}
\newcommand{\taucycl}{\mathrlap{\kern0.42em{\bullet}}\circlearrowright}

\begin{document}
\title{Electromagnetic coupling in tight-binding models for strongly correlated light and matter}

\author{Jiajun Li}
\affiliation{Department of Physics, University of Erlangen-N\"urnberg, 91058 Erlangen, Germany}
\author{Denis Golez}
\affiliation{Center for Computational Quantum Physics, Flatiron Institute, 162 5th Avenue, New York, NY 10010, USA}
\author{Giacomo Mazza}
\affiliation{Department of Quantum Matter Physics, University of Geneva, 
1211 Geneva 4, Switzerland}
\affiliation{CPHT, CNRS, Ecole Polytechnique, IP Paris, F-91128 Palaiseau, France}
\affiliation{Coll\`{e}ge de France, 11 place Marcelin Berthelot, 75005 Paris, France}
\author{Andrew J. Millis}
\affiliation{Center for Computational Quantum Physics, Flatiron Institute, 162 5th Avenue, New York, NY 10010, USA}
\affiliation{Department of Physics, Columbia University 538 West 120th Street, NY NY 10027 USA}
\author{Antoine Georges}
\affiliation{Coll\`{e}ge de France, 11 place Marcelin Berthelot, 75005 Paris, France}
\affiliation{Center for Computational Quantum Physics, Flatiron Institute, 162 5th Avenue, New York, NY 10010, USA}
\affiliation{CPHT, CNRS, Ecole Polytechnique, IP Paris, F-91128 Palaiseau, France}
\affiliation{Department of Quantum Matter Physics, University of Geneva, 
1211 Geneva 4, Switzerland}
\author{Martin Eckstein}
\affiliation{Department of Physics, University of Erlangen-N\"urnberg, 91058 Erlangen, Germany}

\begin{abstract}
We discuss the construction of low-energy tight-binding Hamiltonians for condensed matter systems with a strong coupling to the quantum electromagnetic field. Such Hamiltonians can be obtained by projecting the continuum theory on a given set of Wannier orbitals. However, different representations of the continuum theory lead to different low-energy formulations, 
because different representations may entangle light and matter, transforming orbitals into light-matter hybrid states before the projection. In particular, a multi-center Power-Zienau-Woolley transformation yields a dipolar Hamiltonian which incorporates the light-matter coupling via both Peierls phases and a polarization density. We compare this dipolar gauge Hamiltonian and the straightforward Coulomb gauge Hamiltonian for a one-dimensional solid, to describe sub-cycle light-driven electronic motion in the semiclassical limit, and a coupling of the solid to a quantized cavity mode which renormalizes the band-structure into electron-polariton bands. Both descriptions yield the same result when many bands are taken into account, but the dipolar Hamiltonian is more accurate when the model is restricted to few electronic bands, while the Coulomb Hamiltonian requires fewer electromagnetic modes.
\end{abstract}
\date{\today}

\maketitle

\section{Introduction}

In recent years, many condensed matter experiments have explored the intriguing phenomena which arise when matter is driven by electromagnetic fields far beyond the linear response regime \cite{Basov2017}. On the one hand, this includes highly nonlinear electron dynamics induced by {\em classical} electromagnetic fields,  such as light-driven Bloch oscillations \cite{Schubert2014} and high-harmonic generation, or the engineering of entirely new states, so-called Floquet phases, under strong time-periodic driving \cite{Oka2009,Wang2013,McIver2018}. Even more intriguing proposals for hybrid light-matter states have been made for the case when the {\em quantum} nature of the electromagnetic field becomes relevant \cite{Schachenmayer2015, Kiffner2019, Schlawin2019, Sentef2018b, Mazza2019, Andolina2019, Wang2019, Curtis2019, Kiffner2019b,Orgiu2015}, through structuring the photon modes using a cavity. Cavity quantum electrodynamics (QED) has reached the regime of ultra-strong coupling between few emitters and photons \cite{Frisk-Kockum2019} and demonstrated the possibility to control chemical reactions \cite{Thomas2019}. First experiments with condensed matter systems in cavities have led to tantalizing observations, including an enhancement of the superconducting transition temperature of a material through coupling to vacuum fluctuations \cite{Anoop2019}.

The rich physics of complex condensed matter systems is largely understood in terms of minimal tight-binding models, which describe interacting electron systems on a lattice with only a few valence orbitals per site. The recent developments in cavity QED, therefore, call for tight-binding models which incorporate the electromagnetic field, and can thus provide a low-energy description for the ultra-strong light-matter coupling in solids, complementary to first-principle approaches \cite{Ruggenthaler2014, Flick2018,Schaefer2018,Nielsen2018}. 

However, while the continuum formulation of quantum electrodynamics is textbook knowledge \cite{LoudonBook,TannoudjiBook}, electromagnetic coupling when the electronic hamiltonian is projected to a restricted low-energy model raises subtle issues. The most straightforward approach to derive few level models in atomic physics or few band models in condensed matter is to project the continuum theory on a subset of orbitals.  
However, while the exact theory is invariant under canonical transformations, the accuracy of a projection to a subset of orbitals in general depends on the representation resulting from the choice of canonical variables. For example, in a representation in which the canonical field variable is the macroscopic displacement field rather than the microscopic electric field, the matter orbitals are hybrid light-matter objects, the matter-field coupling appears differently, and the accuracy of a truncation to a small number of orbitals will change.
The differences in the 
projected light-matter Hamiltonians have been recently discussed very actively in relation to the derivation of few level models for individual atoms in cavity QED \cite{Bernardis2018, Bosman2017, Gely2017, Vukics2014,Bernardis2018b, Di-Stefano2019}, motivated in part by long-standing debates on fundamental questions regarding the interpretation of the superradiant phase transition \cite{Dicke1954, Rzaewski1975, Keeling2007}. In the solid, one should expect a similar dependence of the reduced low-energy Hamiltonian on the starting point of the projection.
The Wannier orbital onto which the projection is performed has a different meaning in different 
representations
, and 
one may consider which choice best represents
the physics within a minimal set of bands. 

Tight-binding models with a coupling to the electromagnetic field have a long history in semiclassical description, where the tight-binding Hamiltonians $H[\bm A, \phi]$ is written in terms of the scalar and vector potentials $\bm A(\bm r,t)$ and $\phi(\bm r,t)$. 
The most widely used minimal semiclassical Hamiltonian for electrons with charge $q$ in one band is obtained by the Peierls substitution \cite{Peierls1933},
\begin{align}
\label{jhedvqx}
H=\sum_{\alpha,\alpha'} t_{\alpha,\alpha'} e^{iq\chi_{\alpha,\alpha'}}\,c_{\alpha}^\dagger c_{\alpha'}   + \sum_{\alpha}q\phi_\alpha c_{\alpha}^\dagger c_{\alpha}.
\end{align}
Here $c_{\alpha}^\dagger$ ($c_{\alpha}$) are creation (annihilation) operators for an electron in a Wannier orbital localized at site $\bm R_\alpha$ of a given lattice, $t_{\alpha,\alpha'}$ denotes the tunnelling matrix elements in the absence of electromagnetic fields, and the Peierls phase factors are given in terms of the vector potential by 
\begin{align}
\chi_{\alpha,\alpha'} =  \int_{\bm R_{\alpha'}}^{\bm R_{\alpha}} d\bm r \cdot \bm A(\bm r),
\end{align}
where the integral is taken along a straight line. It must be emphasized that the gauge in Eq.~\eqref{jhedvqx} is not fixed, but the Hamiltonian defines a gauge theory in which the physics remains invariant under the transformation 
\begin{align}
\label{fghjkl01}
\bm A
\to\bm A+\bm \nabla \Lambda,
\,\,
\phi
\to\phi-\partial_t \Lambda,
\,\,\,
c_{\alpha}
\to c_\alpha e^{iq\Lambda(\bm R_\alpha)}, 
\end{align} 
with an arbitrary function $\Lambda(\bm r,t)$. In this aspect, the Peierls Hamiltonian fundamentally differs from any naively projected continuum Hamiltonian, such as a projection of the Coulomb gauge Hamiltonian onto a subset of bands. An elegant way to fix the light-matter coupling matrix elements in more general (multi-band) semiclassical Hamiltonians is in fact to request the existence of such a gauge structure \cite{Boykin2001}. Gauge-invariant semiclassical Hamiltonians like Eq.~\eqref{jhedvqx} turn out to be not only conceptually elegant but also powerful in practice, as strong-field phenomena in solids, such as optically driven Bloch oscillations, can be captured even within a single-band approximation. 

In contrast to the semiclassical description, recent studies regarding the quantum light-matter coupling often rely on a linearized light-matter coupling or on a projection of the Coulomb gauge Hamiltonian on the valence bands. This is certainly valid when the coupling is not too strong, but must be carefully reconsidered in the ultra-strong coupling regime. One cannot easily quantize a semiclassical description by replacing the classical fields $\bm A$ by quantum fields, because the semiclassical approximation misses field-induced interactions and Lamb-shifts in the solid,  but one may impose that the semiclassical approximation to the {\em projected} quantum theory should lead to the known semiclassical tight-binding descriptions. Among the class of quantum light-matter Hamiltonians which are derived in this paper from the continuum theory, 
one particular representation, often referred to in the quantum optics literature as ``the dipolar gauge", 
results in a  light-matter coupling via Peierls phases and inter-band dipolar matrix elements, which has the known semiclassical limit and seems to provide an accurate few-band representation of the physics.  

The article is organized as follows. In Sect.~\ref{sec:222}, we introduce the quantum light-matter Hamiltonian in Coulomb gauge and discuss the general formalism of unitary transformations of the light-matter coupled theory. The multi-center Power-Zienau-Woolley (PZW) transformation is introduced to obtain the general quantum Hamiltonian with Peierls phases and inter-band dipolar matrix elements, 
to which we refer, following the convention, as the dipolar gauge Hamiltonian.  The semiclassical limit of the dipolar gauge Hamiltonian then does indeed both have the gauge invariance given by Eq.~\eqref{fghjkl01}, and it leads to a faster convergence to the full description as the number of electron bands taken into account is increased. In Sec.~\ref{seconeband}, we exemplarily consider in detail the example of a one-dimensional solid, both to analyze the strongly driven semiclassical dynamics, and to evaluate the light-dressed electron polariton band-structure in a cavity. Sect.~\ref{sec:555} provides a conclusion and outlook. 

\section{Tight-binding models in Coulomb and dipolar gauge}
\label{sec:222}

\subsection{Continuum light-matter Hamiltonian in Coulomb gauge}
\label{seckqhjxnla}

In Coulomb gauge, the electromagnetic field is expressed in terms of the transverse vector potential $\bm A(\bm r)$ ($\bm \nabla\cdot\bm A=0$), and its canonical conjugate variable $\bm \Pi(\bm r)$ is related to the electric field (see below) \cite{LoudonBook, TannoudjiBook}.  The Hamiltonian is split in a light and matter contribution as
\begin{align}
\label{oossoo}
H^C&=H_{el}^C+H_{em}.
\end{align}
Here  $H^C$ denotes the minimal coupling Hamiltonian for electrons with charge $q$ in the continuum, 
\begin{align}
H_{el}^C =
\int d^3\bm r\,
\psi_{\bm r}^\dagger
\frac{(-i\bm \nabla -q \bm A(\bm r))^2}{2m}
\psi_{\bm r}
+H_{latt}
+
H_{int}^C,
\label{jjjjxsjsjsd}
\end{align}
where $\psi_{\bm r}^\dagger$ and $\psi_{\bm r}$ are creation and annihilation operators for electrons at point $\bm r$ (the spin index is suppressed throughout the paper for simplicity); $H_{int}^C$ is the instantaneous Coulomb interaction, and  $H_{latt}$ is the lattice potential, which is taken to be a given external potential if the nuclei approximately remain at fixed positions. 

The second part $H_{em}$ of the Hamiltonian \eqref{oossoo} describes the energy of the transverse electromagnetic fields in the empty cavity.
The cavity can be included by allowing for a space-dependent background dielectric function $\epsilon(\bm r)$, which describes a set of dielectric mirrors. Note that $\epsilon(\bm r)$ does not yet include the part of the matter which will be treated explicitly within a microscopic description below. The sole  purpose of the background dielectric is to implement the cavity-induced change of mode density at the position of the solid, and we can therefore adopt the following simple assumptions: (i) The background medium is isotropic and the mirrors are lossless, i.e.,  $\epsilon(\bm r)$ is a frequency independent scalar quantity. (ii) The material of interest is spatially separated from the mirrors, and the background dielectric constant is taken to be $\epsilon(\bm r)=1$ throughout the material.  Quantization of the electromagnetic field inside a linear lossless dielectric medium is usually carried out within a generalized Coulomb gauge $\bm{\nabla}\cdot [\epsilon(\bm{r})\bm{A}]=0$ \cite{Glauber1991}. 
This method has been frequently adopted in the literature and recently used to discuss few-mode approximations in an open cavity \cite{Lentrodt2018}. It formally considers all normal modes of the  overall system-plus-environment (the ``universe"), and thus avoids the subtleties raised due to imposing boundary conditions  for a finite domain and can be used as a starting point of {\it ab initio} description of a lossless cavity. We will follow this procedure below, but because of assumption (ii) , inside the material of interest the usual Coulomb gauge condition holds,  the commutator relations among the electromagnetic fields are the same as in  free space (see below), and the minimal coupling Hamiltonian is given  by Eq.~\eqref{jjjjxsjsjsd}.
The electromagnetic energy of the empty cavity become \cite{Glauber1991}
\begin{align}
H_{em}
&=
\frac{1}{2}
\int d^3\bm r
\Big[
\frac{1}{\epsilon_0\epsilon(\bm r)}\bm \Pi^2+
\frac{1}{\mu_0}(\bm \nabla \times \bm A)^2
\Big].
\label{lwcbas}
\end{align}
The longitudinal fields are not independent dynamical degrees of freedom, but fixed by the charge distribution, and their energy is included in the long-range Coulomb interaction $H^C_{int}$\cite{LoudonBook, TannoudjiBook}.
The effect of the background dielectric constant on the static Coulomb interaction can be assumed being already absorbed in $H^C_{int}$. (Note that, in any case, for almost all model-based calculations in condensed matter the Coulomb interaction matrix elements are approximated, such as through a simple local Hubbard interaction.)

The resulting Heisenberg equations of motion for $\boldsymbol{A}$ and $\boldsymbol{\Pi}$  can be derived using the canonical commutation relation of $\bm A$ and $\bm \Pi$ (Appendix~\ref{AppendixA}) and are the transverse components of the Maxwell equations with the current operator
\footnote{The mathematical meaning of the functional derivative  $\frac{\delta}{\delta \bm A}$  is explained in the appendix.}
\begin{align}
\label{jhqvsv}
\bm j_C(\bm r)
&=
-\frac{\delta H^{C}_{el}}{\delta \bm A(\bm r)},
\end{align}
if the canonical variables are related to the electric and magnetic field as $\bm B (\bm r)= \bm \nabla\times \bm A (\bm r)$
and 
\begin{align}
\label{kcwbslxNAZ}
\bm \Pi(\bm r) = 
-\bm{D}^T(\bm{r}), 
\end{align}
where $\bm{D}(\bm{r})=\epsilon_0\epsilon(\bm{r})\bm{E}(\bm{r})$ is the electric displacement vector
(which includes  polarization effects due to the background medium, but not due to the material of interest.)
Here and in the following, superscripts $T$ and $L$ refer to transverse and longitudinal components of a vector field, respectively.
Note that $\bm \Pi(\bm r) = - \epsilon_0\epsilon(\bm r) \bm E^T(\bm r)$ inside the matter, where $\epsilon(\bm{r})=1$ is assumed. 
The current operator satisfies the continuity equation $\bm \nabla \cdot \bm j  +  \partial_t \rho=0$ with the microscopic charge density 
\begin{align}
\label{sdfghzfghj}
\rho(\bm r)=\psi_{\bm r}^\dagger \psi_{\bm r}.
\end{align}
Note that with Eq.~\eqref{kcwbslxNAZ}, $H_{em}$ takes the standard form $\frac12\int d^3 \bm r [\epsilon \epsilon_0(\bm E^T)^2+\bm B^2/\mu_0] $ for the energy stored in the transverse modes. The longitudinal components of the electric field is constrained by the charge distribution, $\bm\nabla\cdot \bm E^L =\rho-\rho_{\rm background}$. 

In the derivation of low-energy Hamiltonians, one must restrict the Hilbert space to certain energy bands in the solid and certain modes of the electromagnetic field. The selection of modes is an essential part in defining the low energy Hamiltonian, and we adopt the following general notation:

For the matter, one can assume the existence of an electronic single-particle basis of localized orbitals $w_{\alpha}(\bm r)$, each centered around a position $\bm R_\alpha$. Typically, these will be Wannier orbitals $w_{\bm R,n}(\bm r)\equiv w_{n}(\bm r-\bm R)$ 
\footnote{See, for example, C.~Kittel \emph{Introduction to Solid State Physics}, New York: Wiley (1976)}
, where $\alpha=(\bm R,n)$ labels the orbital $n$ and the lattice site 
(with $\bm R_\alpha=\bm R$)
, but in practice we only assume that they are sufficiently localized on the atomic scale and mutually orthogonal 
\begin{align}
&\int d^3\bm r\,w_{\alpha}(\bm r)^* w_{\alpha'}(\bm r) = \delta_{\alpha,\alpha'}.
\end{align}
Because of the orthogonality, the corresponding creation and annihilation operators $c_{\alpha}^\dagger$ and $c_{\alpha}$ for electrons in the field-independent Wannier orbitals satisfy canonical anticommutation relations, and field operators can then be expanded like
\begin{align}
\label{xqQSKWKQK}
\psi_{\bm r}
= \sum_{\alpha} w_{\alpha}(\bm r) c_{\alpha},
\,\,\,\,\,
c_{\alpha}
=
\int d^3\bm r  \,w_{\alpha}(\bm r)^* \psi_{\bm r}.
\end{align}
The expansion of the electromagnetic field into a set of transverse modes is written as
\begin{align}
\label{abkxlnm;z}
&\bm A(\bm r)
=
\sum_{\nu}
\bm {\phi}_\nu(\bm r) Q_{\nu}, 
\\
\label{hhhshwh}
&\bm \Pi(\bm r)
=
\sum_{\nu}
\bm\phi_{\nu}(\bm r)^*
\eta_\nu \epsilon(\bm r)
\,\Pi_\nu,
\end{align}
where the operators $Q_{\nu}$ and $\Pi_\nu$ denote the canonical variables, $[Q_\nu,\Pi_{\nu'}]=i\delta_{\nu,\nu'}$, and the  
factor $\epsilon(\bm r)$ and an arbitrary additional rescaling factor $\eta_\nu $ have been introduced for later convenience.  
To be general, we allow for complex mode functions, which includes propagating modes such as plane waves in free space, and bound states in a cavity which can typically be chosen to be real.
The 
expansion shown above is subject to the reality conditions $\bm{A}^\dag=\bm{A}$ and 
$\bm{\Pi}^\dag=\bm{\Pi}$, leading to, in general, a nontrivial representation of $Q_\nu,\Pi_\nu$ in terms of creation and 
annihilation operators $a_\nu,a^\dag_\nu$. This has been discussed in existing literature \cite{Glauber1991, Lentrodt2018}, 
and is briefly summarized in the appendix \ref{norm_exp}.
The mode functions
are supposed to satisfy the same gauge condition 
as the vector potential, $\bm{\nabla}\cdot [\epsilon(\bm{r})\bm{\phi}_\nu]=0$ 
($\bm{\nabla}\cdot\bm{\phi}_\nu=0$ inside the matter of interest),
and are orthogonal
 (with respect to $\eta_\nu \epsilon(\bm r)$) \begin{align}
&\int d^3\bm r\,
\eta_\nu \epsilon(\bm r)
\, \bm{\phi}_{\nu}(\bm r)^* \cdot \bm {\phi}_{\nu'}(\bm r)
=
\delta_{\nu,\nu'},
\label{inner}
\end{align}
and provide a complete set of transverse functions, with the inverse transformation 
\begin{align}
Q_\nu&=\int d^3\bm r \,\eta_\nu \epsilon(\bm r) \bm A(\bm r)\cdot\bm \phi^*_\nu(\bm r),
\\
\label{hhhshwhinv}
\Pi_\nu&=\int d^3\bm r \,\bm \Pi(\bm r) \cdot\bm\phi_\nu(\bm r).
\end{align}
For example, 
in an extended medium these mode functions can be taken as plane waves.
More generally,
the modes can be taken as normal modes of the resonator Hamiltonian $H_{em}$ [Eq.~\eqref{lwcbas}] so that $H_{em} = \sum_\nu \frac{\omega_\nu}{2} (Q_\nu^2 + \Pi_\nu^2)$,
or from a suitable multi-mode approximation \cite{Lentrodt2018} of an open cavity. Using Maxwell equations for the free cavity and Eq.~\eqref{kcwbslxNAZ}, the diagonalization is achieved by solving the generalized eigenvalue problem
\begin{align}
\label{nm,mnm,m}
\omega_\nu^2\mu_0 \epsilon_0\epsilon(\bm r) \bm b_\nu(\bm r) =\bm{\nabla}\times\bm{\nabla}\times \bm b_\nu(\bm r),
\end{align}
for transverse mode functions  $\bm b_\nu(\bm r)$ that are orthogonal with respect to $\epsilon(\bm r)$,
$\int d^3 \bm r \,\epsilon(\bm r)\bm b_\nu(\bm r) \cdot \bm b_{\nu'}(\bm r) =\delta_{\nu,\nu'}$. The fields can thus be  expanded as \eqref{abkxlnm;z} and \eqref{hhhshwh} with 
generally non-hermitian
$Q_\nu$, $\Pi_\nu$ and 
functions $\bm \phi_\nu(\bm r) \equiv \bm b_\nu(\bm r)/\sqrt{\epsilon_0\omega_\nu}$, $\eta_\nu\equiv \epsilon_0\omega_\nu$.

\subsection{General gauge transformations}
\label{sec:hauge}

Before the projection to a low-energy manifold, one can choose the explicit form of light-matter coupling by performing a 
canonical
transformation. 
The resulting representation
implies a certain definition of the polarization density, and it specifies the definitions of physical fields, such as the macroscopic electric field, in terms of canonical field variables. The 
transformation therefore modifies the form of light-matter coupling \cite{TannoudjiBook,LoudonBook} and can significantly affect the quality of few-bands approximations. 
In the literature these transforma- tions are often referred to as gauge transformations and we will use this terminology in what follows.
In this section, we recapitulate the general formulation of such a gauge transformation in terms of a unitary transformation which mixes light and matter.

 The representation of the Coulomb gauge Hamiltonian in the Wannier basis can be denoted by the formal expression
\begin{align}
H^{C}[c,c^\dagger,\bm\Pi,\bm A]
\equiv
H^{C}_{el}[c,c^\dagger,\bm\Pi,\bm A]
+
H_{em}[\bm\Pi,\bm A],\nonumber
\end{align}
where the square brackets indicate the dependence on the canonical variables. 
As discussed in the last section, in the Coulomb gauge, the canonical field $\boldsymbol{\Pi}$ is identified with the transverse electric field, while the free charge corresponds to the total particle number operator $\rho(\boldsymbol{r})=\psi^\dag_{\boldsymbol{r}}\psi_{\boldsymbol{r}}$, and no bound charge is present. This can be changed by a general unitary transformation $\mathcal{W}$, where the transformed Hamiltonian reads
\begin{align}
\mathcal{W}\, H^{C}
\mathcal{W}^\dagger
&=
H^{C}[\bar c,\bar c^\dagger,\bar {\bm \Pi},\bar {\bm A} ]
\equiv
H^{\mathcal {W}}[c,c^\dagger, {\bm \Pi}, {\bm A} ].
\label{jhvaxkLZ}
\end{align}
Here the expression of the transformed operators $\bar O = \mathcal{W} O \mathcal{W}^\dagger \equiv \bar O [c,c^\dagger,\bm\Pi,\bm A]$, for $O=c,c^\dagger,\bm \Pi,\bm A$ depends on both matter and fields in general, and the last step simply re-expresses the Hamiltonian in terms of $c,c^\dagger, {\bm \Pi}, {\bm A}$, so that $H^{C}$ and $H^{\mathcal {W}}$ have a different functional dependence on the canonical variables in general.

We emphasize that both in Coulomb gauge and in the $\mathcal{W}$ gauge, the symbols  $c,c^\dagger, {\bm \Pi}, {\bm A}$ will be used to denote a set of operators which satisfy the canonical (anti)-commutation relations, and thus  serve to construct the Hilbert space. In contrast, gauge-invariant physical observables such as the microscopic electromagnetic fields $\bm E$ and $\bm B$ itself do not depend on the gauge, but their representation in terms of the canonical variables does, which will be denoted by a subscript $\mathcal{W}$ or $C$. (We will also introduce a set of operators which are defined differently for each gauge, such as the polarization.) If $X$ is a gauge-invariant observable, its representation in $\mathcal{W}$ gauge is obtained  by
$\mathcal{W}
X_C\mathcal{W}^\dagger
=
X_C[\bar c,\bar c^\dagger, \bar {\bm \Pi},\bar  {\bm A}]
\equiv
X_\mathcal{W}[c,c^\dagger, {\bm \Pi}, {\bm A}]$ (just as the transformation of $H$).
Written for the microscopic fields, the above discussion implies that the canonical variables $c,c^\dagger, {\bm \Pi}, {\bm A}$ correspond to different physical quantities in different gauges. 
For example, we shall see that $\bm \Pi$ in the new gauge will include a contribution from the electric polarization defined by the gauge transformation.

For the discussion of the solid, it is useful to consider a rather general class of gauge transformation which mix light and matter in the form of a linear mapping of the matter operators,
\begin{align}
\label{lcnaxs}
\bar c_{\alpha}[c,\bm A]
\equiv
\mathcal{W}
c_{\alpha}
\mathcal{W}^\dagger
=
\sum_{\alpha'}
W[\bm A]_{\alpha,\alpha'} c_{\alpha'}.
\end{align}
Here  the matrix $W[\bm A]$ depends on the field operator $\bm A$ only (not on $\bm \Pi$), and it is unitary in terms of the matter indices.
One can show that this relation for the electron operators already fixes the transformation of the electromagnetic fields: Because the transformation depends only on $\bm A$, one has $ \mathcal{W}  {\bm A} \mathcal{W}^\dagger = {\bm A}$. Furthermore, the transformation of $\bm \Pi$ can always be represented as a shift by the transverse component of a field $\bm P_ \mathcal{W} (\bm r)$
\begin{align}
\label{ergbnm01}
 \mathcal{W}  {\bm \Pi} \mathcal{W}^\dagger &= {\bm \Pi} + \bm P^T_{\mathcal{W}}(\bm r),
\end{align}
which will be identified as polarization density. In Appendix \ref{sechdjd} we show that the polarization density that corresponds to the general transformation \eqref{lcnaxs} is a simple quadratic form in the matter operators,
\begin{align}
\label{jgkxalsz;}
\bm P^T_{\mathcal{W}}(\bm r)
&=
\sum_{\alpha,\alpha'}
\bm M(\bm r)_{\alpha,\alpha'}^T
\,c_{\alpha}^\dagger
c_{\alpha'},
\\
\label{jgkxalsz;01}
\bm M(\bm r)^T
&=
-iW[\bm A]^\dagger \Big(\frac{\delta}{\delta \bm A(\bm r)}W[\bm A]\Big)^T,
\end{align}
where the derivative with respect to $\bm A$ is understood in terms of the commutator, see Eq.~\eqref{cwjlXA} in App.~\ref{AppendixA}.
By transforming $\bm \Pi$ in the free field Hamitonian $H_{em}$ to $\bm \Pi+\bm P_{\mathcal{W}}^T$, one then arrives at a transformed Hamiltonian which contains both a new light-matter interaction $\sim \bm P\cdot\bm \Pi$, and an induced interaction $\sim \bm P\cdot \bm P$. In summary, the Hamiltonian can be written as 
\begin{align}
H_{\mathcal{W}}=H_{el,\mathcal{W}} + H_{em} + H_{PP} + H_{EP},
\end{align}
where $H_{el,\mathcal{W}}$ is obtained by applying the transformation \eqref{jhvaxkLZ} and \eqref{lcnaxs} to $H_{C,el}$, 
and 
\begin{align}
H_{EP}
&=
\frac{1}{2}\int d^3\bm r
\frac{1}{\epsilon_0\epsilon(\bm r)}
\big[\bm \Pi(\bm r)\cdot \bm P^T_{\mathcal{W}}(\bm r)+h.c.\big],
\label{jlxlxwxww01}
\\
\label{jlxlxwxww}
H_{PP}
&=
\frac{1}{2}\int d^3\bm r
\frac{1}{\epsilon_0\epsilon(\bm r)}\bm P^T_{\mathcal{W}}(\bm r)^2.
\end{align}
(In the second line, $\bm \Pi$ and $\bm P^T_{\mathcal{W}}$ do not commute because $\bm P^T_{\mathcal{W}}$ can depend on the vector potential.) Note that, for all matter quantities, such as $H_{PP}$, the $\epsilon(\bm r)$ can essentially be discarded under the assumption $\epsilon(\bm r)=1$ inside the matter. 

With the Hamiltonian $H_{\mathcal{W}}$, Heisenberg equations can be identified with the {\em macroscopic} Maxwell equation for the transverse fields with the current density 
\begin{align}
\label{jhqvsv01}
\bm J_{\mathcal{W}}
=
-\frac{\delta (H_{el,\mathcal{W}}+H_{EP}+H_{PP})}{\delta \bm A(\bm r)},
\end{align}
if $\bm \Pi$ is identified with a displacement field $\bm D^T_\mathcal{W}$,
\begin{align}
\label{fffafaffa}
&\bm \Pi(\bm r) =
-\bm{D}^T(\bm r)
- \bm P^T_{\mathcal{W}}(\bm r)  \equiv -\bm D^T_{\mathcal{W}}(\bm r),
\end{align}
again with $\bm D^T=\epsilon_0\epsilon(\bm r)\bm E^T$ if $\epsilon(\bm r)$ is constant in the relevant region. 
The current $\bm J_\mathcal{W}$ satisfies the continuity equation with a charge density
\begin{align}
\label{hhhahha}
\rho_{macr,\mathcal{W}}
=
\rho_\mathcal{W}
+
\bm \nabla \cdot \bm P_\mathcal{W},
\end{align}
which we term ``macroscopic charge density'', because the related current is the source term in the macroscopic Maxwell equations. 
Comparison with the microscopic continuity equation shows that $\bm J_{\mathcal{W}}$ is related to the microscopic current by
\begin{align}
\label{cghghvgggs}
\bm j_\mathcal{W} =  \bm J_{\mathcal{W}}  +  \partial_t  \bm P_\mathcal{W},
\end{align}
showing again the gauge-dependent separation of the charge into a macroscopic charge density and a polarization charge $-\bm \nabla \cdot \bm P_\mathcal{W}$.

\subsubsection*{Semiclassical approximation}

The semiclassical approximation corresponds to replacing the electromagnetic field by its expectation value, while leaving the quantum description of the matter. 
Since $\bm{\Pi}$ in the new gauge is a mixed object of light and matter, one needs to make sure the semiclassical limit is taken only for 
the \emph{pure} photon degree of freedom, or the combination $\bm{\Pi} +  \bm P^T_{\mathcal{W}}(\bm r)=-\bm{D}^T$. Specifically, it
 can be obtained by decoupling the square of the operator $\bm \Pi + \bm P^T_{\mathcal{W}}(\bm r)$ in the light-matter Hamiltonian, analogous to a mean-field decoupling $AB\to A\langle B\rangle+\langle A\rangle B - \langle A\rangle\langle B\rangle$ of the products $\bm \Pi\cdot \bm P_\mathcal{W}$ and $\bm P_\mathcal{W}^2$ in  Eqs.~\eqref{jlxlxwxww01} and  \eqref{jlxlxwxww}. The resulting equations, written in terms of the microscopic fields $\bm B(\bm r,t)$ and $\bm E(\bm r,t)$,  are the classical 
Maxwell equations for the transverse fields with the microscopic current $\langle \bm j\rangle$ and charge as a source term. Matter is described by the semiclassical Hamiltonian, 
\begin{align}
H_{sc,\mathcal{W}}=H_{el,\mathcal{W}}[c^\dagger,c,\bm A(\bm r,t)] -\int d^3\bm r
 \bm E^T(\bm r,t) 
\cdot \bm P^T_{\mathcal{W}}(\bm r),
\label{jlxlxwxww03}
\end{align}
up to the constant term $\propto \int \langle \bm P^T_{\mathcal{W}}(\bm r)\rangle^2$. 
The dielectric constant $\epsilon(\bm{r})$ is assumed to be 
uniform inside the matter and $\bm{D}^T=\epsilon_0\epsilon(\bm{r})\bm{E}^T$.
 In these equations, the vector potential is still transverse, $\bm E^T(\bm r,t)=-\partial_t \bm A(\bm r,t)$. The current is given by Eq.~\eqref{cghghvgggs}, with the 
macroscopic charge contribution Eq.~\eqref{jhqvsv01}, $\bm J_{sc,\mathcal{W}}= -\frac{\delta H_{sc,\mathcal{W}}}{\delta \bm A(\bm r)}$, and the 
polarization charge contribution $\partial_t \langle\bm P_{\mathcal{W}}(\bm r)\rangle$. 

One possible requirement for the construction of tight-binding light-matter Hamiltonians is to find a gauge transformation such that the semi-classical Hamiltonian after projection to subset of bands has an explicit gauge structure as defined by Eq.~\eqref{fghjkl01}.

\subsubsection*{Restriction of the cavity modes}

Often it is useful to express the fields using a general mode expansion \eqref{abkxlnm;z} and \eqref{hhhshwh}.  With the replacement \footnote{Note that a dependence on $\bm A$ implies a dependence on all $Q_\nu$, and the functional derivative Eq.~\eqref{cwjlXA} becomes $[\mathcal{O}[\bm A],\Pi_\nu]=i\Big(\frac{\delta \mathcal{O}[\bm A]}{\delta Q_\nu}\Big)^T$. 
Again, the functional derivative is, indeed, defined by the 
corresponding commutator.}
\begin{align}
\label{fghjfgfhhgh}
\frac{\delta }{\delta \bm A(\bm r)}
=
\sum_{\nu}
\eta_\nu \epsilon(\bm r)
\bm\phi_{\nu}^*(\bm r)
\frac{\delta }{\delta Q_\nu}
\end{align}
in Eq.~\eqref{jgkxalsz;01}, we obtain the expansions of the polarization density $\bm P^T_\mathcal{W}(\bm r)$ in terms of the mode functions $\bm \phi$,
\begin{align}
\label{kvxlnzA;}
\bm P^T_{\mathcal{W}}(\bm r)
&=
\sum_{\nu}
\eta_\nu \epsilon(\bm r)
\bm\phi_{\nu}^*(\bm r)
P^T_{\mathcal{W},\nu},
\\
\label{kvxlnzA03}
P^T_{\mathcal{W},\nu}
&=
\sum_{\alpha,\alpha'}
c_{\alpha}^\dagger 
(M_\nu)_{\alpha,\alpha'}
c_{\alpha'},
\\
\label{fghjkljh}
M_\nu
&=
-i
W[\bm A]^\dagger
\frac{\delta }{\delta Q_\nu}
W[\bm A].
\end{align}
The current operator \eqref{jhqvsv01} is expanded in an analogous manner. Within this expansion, the light-matter Hamiltonian becomes
\begin{align}
\label{jbvbjbx01}
H_{EP}
&=\frac{1}{2\epsilon_0}
\sum_{\nu}\eta_\nu
\big(\Pi_\nu^\dagger P_{\mathcal{W},\nu}^{T} + h.c.\big),
\\
\label{jbvbjbx02}
H_{PP}
&=\frac{1}{2\epsilon_0}
\sum_{\nu}\eta_\nu
(P_{\mathcal{W},\nu}^{T})^\dagger P_{\mathcal{W},\nu}^{T}.
\end{align}

This equation is particularly useful when the number of modes is restricted. For example, a coarse graining of the fields can formally be achieved by truncating the  relevant modes $\nu$ to a low energy subspace, i.e., introducing a momentum cutoff, or one can restrict the modes to few normal modes of the cavity resonator or a suitable multi-mode approximation \cite{Lentrodt2018}. 
This truncation must consistently treat all degenerate modes for a frequency $\omega_\nu$, so that the hermiticity of the dynamical 
variables $\bm A$ and $\bm \Pi$ is guaranteed (see appendix~\ref{norm_exp} for more details).
It is important to note that such a truncation changes the dipolar interaction $H_{PP}$ and the light-matter coupling $H_{EP}$ in a consistent manner. For example, a restriction to a single mode which is homogeneous over the solid is consistent with an all to all interaction $\propto \bar P^2$, where $\bar P$ is a volume averaged polarization (see examples in Sec.~\ref{seconeband}).

\subsection{PZW transformation and dipolar gauge}

For the description of a single atom in a cavity at strong coupling, one often uses the PZW transformation to change the light-matter coupling from the form $\boldsymbol{A}\cdot\boldsymbol{p}$ to $\boldsymbol{E}\cdot\boldsymbol{r}$,  which is suitable when electrons are localized close to an atomic center ($\bm r=\bm 0$). In second quantization, the unitary transformation reads
\begin{align}
\label{,acbskcslc}
\mathcal{W}_\text{PZW}
&=
\exp\Big(-iq\int d^3
\bm r
\, \chi(\bm r,\bm 0)\,
\psi_{\bm r}^\dagger\psi_{\bm r}
\Big),
\end{align}
where $\chi(\bm r,\bm r')$ is the line integral over the vector potential along a straight path,
\begin{align}
\label{jjwjwkqblq}
\chi(\bm r,\bm r')
&=
\int_{\bm r'}^{\bm r} d\bm s \cdot \bm A(\bm s).
\end{align}
The PZW transformation is particularly useful as a starting point for the multipolar expansion of the atom-field interaction, where an electron remains localized close to a given atomic center. In the solid, the choice of a fixed origin is  however not very convenient, as it explicitly breaks the spatial translational invariance. For the derivation of the semiclassical Peierls Hamiltonian \eqref{jhedvqx}, Luttinger introduced a similar {\em multi-center PZW transformation} \cite{Luttinger1951}. The analog for the quantum case is the definition of field-dependent hybrid light-matter orbitals 
\begin{align}
\label{sbxalgg}
\tilde w_{\alpha}(\bm r)
=
e^{-iq\chi(\bm r,\bm R_\alpha)} w_{\alpha}(\bm r),
\end{align}
where the phase due to the vector potential for each orbital is defined relative to the center of the Wannier orbital. However, these orbitals are not orthogonal. It is easy to see that the overlap matrix instead reads
\begin{multline}
\label{ghgsgw}
\int d^3\bm r\,\tilde w_{\alpha}(\bm r)^\dagger \tilde  w_{\alpha'}(\bm r) 
\\=
e^{-iq\chi_{\alpha,\alpha'}} 
\int d^3\bm r\,w_{\alpha}(\bm r)^*
e^{iq\Phi(\bm R_{\alpha'},\bm r,\bm R_\alpha)} 
w_{\alpha'}(\bm r),
\end{multline}
where $\Phi(\bm R_{\alpha'},\bm r,\bm R_\alpha)$ is the magnetic flux through the oriented triangle $\bm R_{\alpha'}\to\bm r\to\bm R_{\alpha}\to\bm R_{\alpha'}$, and we have used the shorthand notation for the Peierls phase $\chi_{\alpha,\alpha'}=\chi(\bm R_\alpha,\bm R_{\alpha'})$.  
Because the Wannier orbitals are exponentially localized, $\Phi $ is of the order of the magnetic flux per lattice plaquette, and the corresponding deviation of the overlap matrix of order $q\Phi/\hbar$ is typically much smaller than one. For example, for classical electromagnetic waves in vacuum, even an almost atomically strong electric field amplitude $1~$MV/cm implies only a magnetic field of order $0.3$ Tesla, and that the flux $\Phi_0$ through a plaquette of size $10^{-19}m^2$  gives  $e\Phi_0/\hbar \sim10^{-4}$. One could re-orthogonalize the field-dependent orbitals order by order in the magnetic flux, which would finally lead to a light-matter Hamiltonian including an explicit interaction with the magnetic field (magnetic dipolar interactions). While this is possible, in the present manuscript we neglect all magnetic field-dependent matrix elements of that order (magnetic flux per plaquette), thereby obtaining a Hamiltonian containing only the dominant electric dipolar terms (``electric dipole approximation''). 
In practice, one may want to justify the approximation by self-consistently checking the spatial variation of computed field variables.

From now on, we therefore assume that the overlap \eqref{ghgsgw} is given by the identity $\delta_{\alpha,\alpha'}$. The orthogonality implies that one can construct  annihilation  (creation) operators $\bar c_{\alpha}=\int d^3\bm r  \,\tilde w_{\alpha}(\bm r)^* \psi_{\bm r}$ and ($\bar c_{\alpha}^\dagger$) for electrons in the hybrid orbitals which satisfy canonical anti-commutation relations. The transformation from $c$ to $\bar c$ is therefore unitary and of the type \eqref{lcnaxs}, and all properties derived in the previous section apply. In particular, we can directly determine the corresponding Hamiltonian $H_{\rm Dip}$, and the polarization operator $\bm P_{\rm Dip}$. (The subscript refers to ``dipolar'' or PZW gauge.)  For better readability, we have shifted a rather straightforward derivation of the polarization operator and the Hamiltonian in dipolar gauge to  the Appendix \ref{pppqqqppp01} and \ref{pppqqqppp02}, and summarize the results here. 

All derivations  use the approximation that the vector potential varies weakly on the atomic scale, which is consistent with the electric approximation above and corresponds to neglecting magnetic dipolar and electric quadrupolar matrix elements.  We do not make, however, the approximation that the fields vary little over the full crystal (i.e., treating the crystal as a big molecule), as this would neglect the momentum-dependence of the field modes inside the solid from the outset. The slow variation of the fields is usually justified by an energy separation. In the mode expansion \eqref{abkxlnm;z} of the electromagnetic fields, modes which vary on some short scale $\lambda_c$ can be disregarded because the corresponding energy is large compared to the relevant electronic transitions in the solid (coarse graining). In the solid, $\lambda_c$ could be a UV wavelength such that $\hbar\omega_c=2\pi\hbar c/\lambda_c$ is the order of several $eV$, but still $\lambda_c$ spans many lattice spacings. 

For later use, it is convenient to represent both the Hamiltonian and the polarization using the mode expansion  \eqref{abkxlnm;z} and \eqref{hhhshwh} rather than the continuum, as in Eqs.~\eqref{jbvbjbx01} and \eqref{jbvbjbx02}. For the expansion coefficients \eqref{kvxlnzA03} we get
\begin{align}
\label{gegege}
P_\mathcal{{\rm Dip},\nu}
&=
q
\sum_{\alpha,\alpha'}
c_{\alpha}^\dagger c_{\alpha'}
\tilde D^\nu_{\alpha,\alpha'},
\end{align}
where we introduced the dipolar matrix elements
\begin{align}
\label{cghjhbjknjkjn0}
\bm D_{\alpha,\alpha'}
&=
\int d^3{\bm r}\,
w_{\alpha}(\bm r)^*
\,\bm r\,
w_{\alpha'}(\bm r),
\end{align} 
which are then projected on the mode functions and dressed by a Peierls phase $\chi_{\alpha,\alpha'}$,
$\tilde D^{\nu}_{\alpha,\alpha'}=\big(
\bm \phi_\nu(\bm R_{\alpha,\alpha'})
\cdot
\bm D_{\alpha,\alpha'}\big)e^{iq\chi_{\alpha,\alpha'}}$; $\bm R_{\alpha,\alpha'}=(\bm R_{\alpha,}+\bm R_{\alpha'})/2$ is the position of the bond $(\alpha,\alpha')$. 
Note that, the matrix $\bm D_{\alpha\alpha'}$ in (42) only contains the dipole contribution, due to the additional assumption that the mode functions vary slowly within the unit cell (see appendix). In general, the derivation of the polarization operator \eqref{kvxlnzA03} within the multi-center PZW transformation can be systematically extended by expanding the spatial dependence of the fields within the unit cell; this would yield the contributions from quadrupolar and even higher-order matrix elements to the field $\bm P$.
The Hamiltonian is given by (see App.~\ref{pppqqqppp02})
\begin{align}
\label{fghjfghjkfghj01}
H_{\rm Dip} = H_{em} + H_{{\rm Dip},el} + H_{EP}+H_{PP}.
\end{align}
Here $H_{{\rm Dip},el}$ is obtained by dressing all matrix elements in the field free Hamiltonian $H_{el,C}[\bm A=0]$ with Peierls factors, i.e., an operator $O_C=c_{\alpha}^\dagger O_{\alpha,\alpha'} c_\alpha$ becomes $O_{\rm Dip}=c_{\alpha}^\dagger e^{iq\chi_{\alpha,\alpha'}}O_{\alpha,\alpha'} c_\alpha$. This replacement holds both for single-particle terms and for two-particle terms (interactions) which can be written as products of such operators.
The dipolar light-matter interaction $H_{EP}$ is obtained by inserting the polarization operator \eqref{gegege} into the general expression \eqref{jbvbjbx01},
\begin{align}
\label{fghjfghjkfghj02}
H_{EP}
&=
\sum_{\nu}
\frac{q\eta_\nu}{2\epsilon_0}
\sum_{\alpha,\alpha'}
\Big(
\Pi_\nu^\dagger
c_{\alpha}^\dagger c_{\alpha'} \tilde D^{\nu}_{\alpha,\alpha'}+h.c.\big).
\end{align}
Analogously, we obtain the $\bm P^2$ term \eqref{jbvbjbx02},
\begin{multline}
\label{fghjfghjkfghj03}
H_{PP}
=
\sum_{\nu}
\frac{q^2\eta_\nu}{2\epsilon_0}
\sum_{\alpha,\alpha'}
c_{\alpha}^\dagger c_{\alpha'}
\big(\tilde D^{\nu}(\tilde D^{\nu})^\dagger\big)_{\alpha,\alpha'}
\\
+
\sum_{\nu}
\frac{q^2\eta_\nu}{2\epsilon_0}
\sum_{\alpha,\alpha',\beta,\beta'}
c_{\alpha}^\dagger 
c_{\beta'}^\dagger 
c_{\beta}
c_{\alpha'}
\tilde D^{\nu}_{\alpha,\alpha'}
(\tilde D^{\nu})^\dagger_{\beta',\beta}.
\end{multline}
We have  written $H_{PP}$ in normal-ordered form, in order to indicate that the first term is a renormalization of a band structure which is relevant even for the case of a single electron.

We remark that the dipolar Hamiltonian has a  gauge structure which makes it invariant under a shift $\bm A\to\bm A+\bm \nabla \Lambda$ and a simultaneous transformation $c_\alpha\to c_\alpha e^{iq\Lambda(\bm R_\alpha)}$. This implies that the current \eqref{jhqvsv01} satisfies the continuity equation with the charge density 
\begin{align}
\rho_{macr,{\rm Dip}}(\bm r)
=
q\sum_{\alpha}
\delta(\bm r-\bm R_\alpha)
c_{\alpha}^\dagger
c_{\alpha}.
\end{align}
This clarifies the separation of the macroscopic and polarization charges according to Eq.~\eqref{hhhahha} in the dipolar gauge.

Finally, the semiclassical approximation \eqref{jlxlxwxww03} in the dipolar gauge becomes
\begin{multline}
H_{sc,{\rm Dip}}=H_{el,{\rm Dip}}[c^\dagger,c,\bm A(\bm r,t)] 
\\
-
q
\sum_{\alpha,\alpha'}
\bm E(\bm R_{\alpha},t)\cdot \bm D_{\alpha,\alpha'}
e^{iq\chi_{\alpha,\alpha'}(t)}
\,
c_{\alpha}^\dagger c_{\alpha'}
\Big],
\label{jlxlxwxww03ggg}
\end{multline}
where the Peierls factors $\chi_{\alpha,\alpha'}(t)$ are calculated using the field $\bm A(\bm r,t)$.  We note that, when a scalar potential term $H_\phi=q\int d^3\bm r \phi(\bm r) \rho_{macr,{\rm Dip}}(\bm r)=q\sum_{\alpha} c_{\alpha}^\dagger c_\alpha \phi(\bm R_\alpha)$ is added, the semiclassical Hamiltonian \eqref{jlxlxwxww03ggg} is invariant under the gauge transformation Eq.~\eqref{fghjkl01}, because  the vector potential enters only via the Peierls phases of the hopping and inter-band dipolar matrices. Since the gauge shift in Eq.~\eqref{fghjkl01} does not mix operators from different Wannier orbitals, the gauge structure is preserved even after projection onto a subset of orbitals. This shows that the semiclassical limit of the dipolar Hamiltonian, even after truncation, falls into the class of models which have a simple gauge structure \cite{Boykin2001}, as defined by Eq.~\eqref{fghjkl01}.

Finally, we note that for general photon modes beyond the dipole approximation, the translational invariance is only recovered for the 
full light-matter hamiltonian, but not individually for the electronic part. 
This should be expected because, obviously, the momentum conservation of electrons is broken due to scattering with
 photons carrying finite momenta. However, the multi-center PZW transformation is still favored in this case, since it leads to the translationally 
 invariant definition of the electric polarization, which is consistent with the general theory in solid-state physics.

\section{One-dimensional solid}
\label{seconeband}

\begin{figure}
 \centerline{\includegraphics[width=0.22\textwidth]{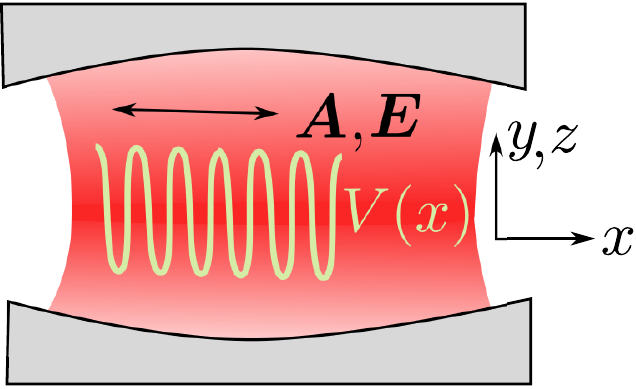}}
\caption{
The setup studied in Sec.~\ref{seconeband}. An electron is subject to a periodic potential $V(x)$ along the $x$ direction and confined to $z=y=0$. In the quantum case, we take into account only one cavity mode with constant amplitude and polarization along the chain. In the classical case, the electron is driven by a time-dependent field $\bm E(t)$, again polarized along the chain.}
\label{figcavity}
\end{figure}

In this section, we will systematically compare the convergence of few-band approximations in the Coulomb and dipolar gauge by numerically solving a specific model. The example we choose is a one-dimensional solid, driven by quantum and classical fields which are polarized along the direction of the solid (Fig.~\ref{figcavity}). Driving the electrons in the system with strong {\em classical} fields leads to phenomena such as nonlinear Bloch oscillations and dynamical localization (band narrowing), which are in part captured already in a suitable single-band model and the Peierls substitution. These effects should have an analog when the solid is strongly coupled to a quantized cavity mode with polarization along the material, which hybridizes with the electronic bands to form electron-polariton bands. 

For simplicity, the discussion is restricted to a single electron, deferring the more involved case of induced interactions to subsequent work. The potential is taken to be sinusoidal, with periodicity $a$ along the direction ($x$) of the solid. The field-free electronic continuum Hamiltonian \eqref{jjjjxsjsjsd} then reads (in first quantization)
\begin{align}
\label{modelAAA}
H_{el}=\frac{p_x^2}{2}+2V_0\cos\left(Gx\right),
\end{align}
where $G=2\pi$, and $V_0$ sets the periodic potential along the solid. Here we have set $\hbar=1$, $a=1$, and $m=1$ (the electron mass), so that length is measured in units of $a$ and energy in units of  $\frac{\hbar^2}{ma^2}$. Eigenstates of the un-driven system are Bloch bands $\langle x|k,m\rangle=\phi_{k,m}(x)$ with quasi-momentum $k\in[-\pi,\pi)$ and band energy $\epsilon_{k,m}$, and  $m=0,1,2,...$. The band structure and dipolar matrix elements are obtained determined using a plane wave representation of the Bloch states (for details, see Appendix \ref{app:fghjkl}). 
For convenience, we will represent all operators in the Bloch basis for both semi-classical and quantum cases.
The Bloch state representation  is complete for a selected subset of bands and therefore equivalent to the Wannier representation used in the general derivation.

\subsection{Semiclassical case: Nonlinear Bloch oscillations}

\subsubsection{Formulation}

\begin{figure*}
 \includegraphics[width=0.99\textwidth]{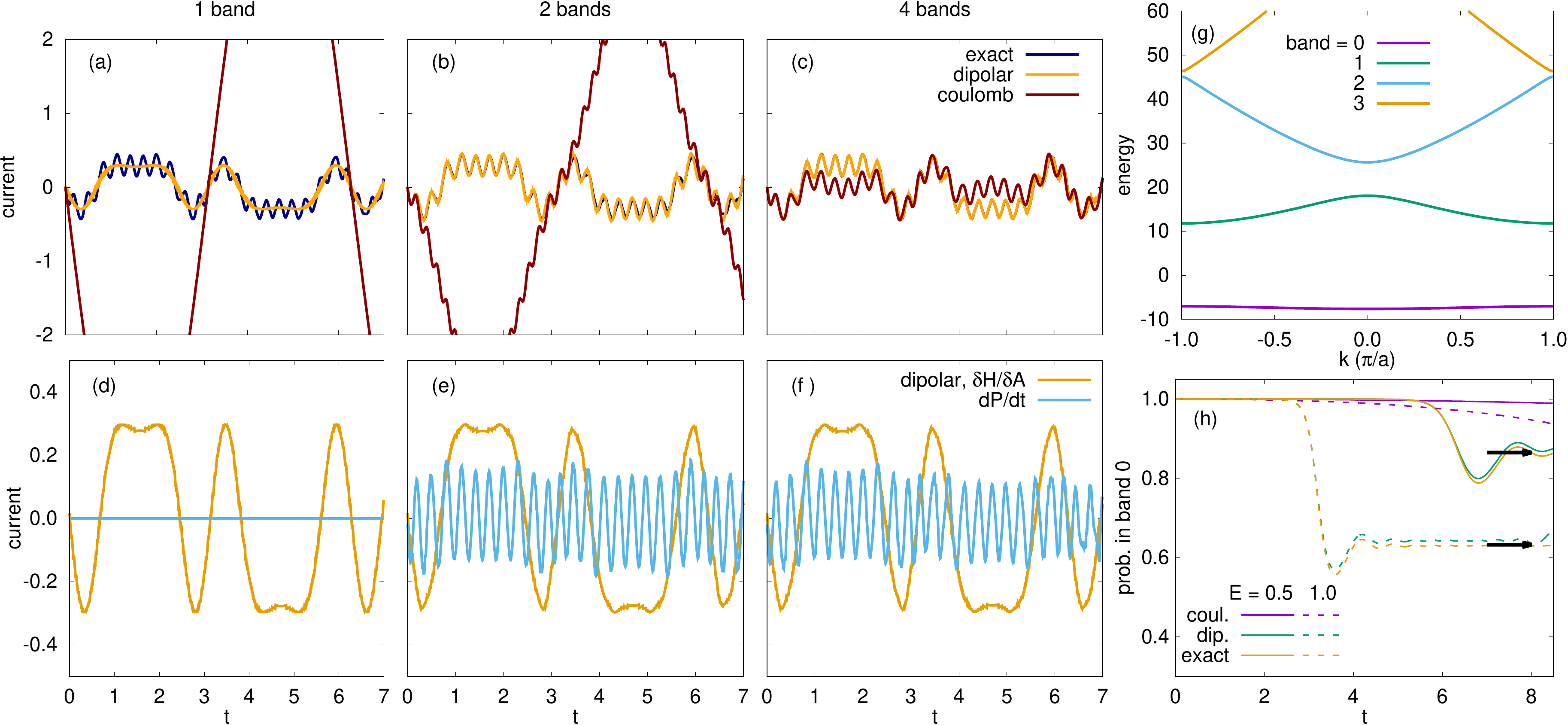}
\caption{Nonlinear Bloch oscillation under a periodic electric field for $V_0=10.0,qA_0=5.0,\omega=1.0$, $q=1$. g) Band structure in the undriven case. a-c) Comparison of the current in the exact simulation with approximate calculations using the Coulomb and dipolar gauge with 1, 2, and 4 bands. d)-f) The two components of the current in dipolar gauge: $J=-\delta H / \delta A$ gives the free-charge current, which is already well reproduced in the 1-band approximation, and $dP/dt$ is the bound-charge current. h) The evolution of probability in the first energy band under a dc-electric field using a two-band cutoff. The black arrows label the probability after one-shot Landau-Zener tunnelling predicted by $1-\exp(-E_g^2/4E)$ with a gap $E_g\approx2$ in the case of $V_0=1.0$.}
\label{fig_curr}
\end{figure*}

We first discuss the description of light-driven dynamics in classical laser fields. The vector potential $A(t)$ is taken to be homogeneous over the solid (i.e., the laser spot extends over many lattice spacings), corresponding to an electric field $E(t)=-\partial_t A(t)$. Only the polarization $A\equiv A_x$ along the solid is considered. One thus must solve for the electron dynamics using the time-dependent semiclassical Hamiltonian
\begin{align}
H(t)=\frac{\left(p_x-qA(t)\right)^2}{2}+2V_0\cos\left(Gx\right).
\end{align}
The exact time-dependent calculation can be carried out straightforwardly in the plane-wave basis. To implement the few-band cutoff in the Coulomb gauge, we  expand the Hamiltonian in the Bloch basis, $\langle k,m | H(t) | k',m'\rangle=\delta_{kk'}H^{\rm C}_{k;m,m'}(t)$,  and neglect all bands with $m>m_{\rm max}$. The matrix elements are given by
\begin{align}
H^{\rm C}_{k;m,m'}&=
\Big[\epsilon_{k;m}+\frac{q^2A(t)^2}{2}\Big]\delta_{m,m'}-qA(t)p_{k;m,m'}, 
\end{align}
with the matrix element  $p_{k;m,m'}=-i\langle \phi_{k,m}|\partial_x|\phi_{k,m'}\rangle$ of the bare momentum operator. The dipolar gauge is directly obtained by projecting the semiclassical Hamiltonian \eqref{jlxlxwxww03ggg}  to a  Bloch-basis. When going from the Wannier representation of Eq.~\eqref{jlxlxwxww03ggg} to the Bloch representation, the single-particle Hamiltonian and  the dipolar matrix $D_{\alpha,\alpha'}\to \delta_{k,k'} D_{k;m,m'}$ become diagonal in momentum (see Appendix for the evaluation of $D$), and the multiplication with the phase factor $\chi_{\alpha,\alpha'}$ corresponds to the Peierls substitution $k\to k-qA(t)$. One finally arrives at  $\langle k,m | H^{\rm Dip}| k',m'\rangle=\delta_{kk'}H^{\rm Dip}_{k;m,m'}$, with 
\begin{align}
\label{ghjkl}
H^{\rm Dip}_{k;m,m'}=
\epsilon_{k-qA(t);m}\delta_{m,m'}-qE(t)D_{k-qA(t);m,m'}.
\end{align}
The exact current operator is 
\begin{align}
\label{cvkbjkjb}
j_x=-\frac{\delta H}{\delta A} = q(p_x -qA(t)).
\end{align}
In Coulomb gauge, its matrix elements read $\langle k,m|j_x|k',m'\rangle = \delta_{kk'} q (p_{k;m,m'}-qA(t)\delta_{m,m'})$. In dipolar gauge, the current has two components, see Eq.~\eqref{cghghvgggs}. One is the macroscopic current, $\langle k,m | J^{\rm Dip} |k',m'\rangle =-\delta_{k,k'} \,\delta H^{\rm Dip}_{k;m,m'} /\delta A = q\,\delta_{k,k'}\,\partial_k H^{\rm Dip}_{k;m,m'}$, which would enter as a source term in the macroscopic Maxwell equations, and the other one is the current of the bound charges, given by the time-derivative of the local polarization $dP^{\rm Dip}/dt$. The polarization operator $P$ has matrix elements  $\langle k,m|P^{\rm Dip}|k',m'\rangle = \delta_{kk'}qD_{k-qA(t),m,m'}$. The current and polarization density are understood to be coarse grained over the solid. 

\subsubsection{Results}

To compare the exact results with the few-band approximations in different gauges, we prepare an initial state with $k=0,m=0$ in all three situations and apply an oscillating vector potential $A(t)=A_0\sin(\omega t)$. The resulting currents are compared in Fig.~\ref{fig_curr}a-c for different band numbers $m_{\rm max}$. We choose $qA_0=5.0$ and $\omega=1.0$. The potential is $V_0=10$, which corresponds to well-separated bands, see Fig.~\ref{fig_curr}g. 

In the exact solution, the current exhibits multiple oscillation periods: On the timescale $2\pi$, which corresponds to the period of $A(t)$, the current has a non-sinusoidal time-dependence (non-linear Bloch oscillations), in addition to fast oscillations which have a frequency of the order of the band splitting. In dipolar gauge, it is the free charge  current $J^{\rm Dip}$ (yellow solid lines in Fig.~\ref{fig_curr}d-f) which gives rise to these nonlinear Bloch oscillations: Due to the large amplitude $A_0$, the mechanical momentum $k-qA(t)$ in dipolar gauge passes the boundary of the first Brillouin zone $[-\pi,\pi)$ during the time-evolution, leading to a reversal of the current. The fast oscillation in the current is instead attributed to the bound-charge current $dP^{\rm Dip}/dt$ (see Fig.~\ref{fig_curr}d-f). The free charge current is converged even in the one-band approximation, while the description of the inter-band polarization current $dP^{\rm Dip}/dt$ is reasonably approximated with a minimal two-band cutoff. In Coulomb gauge, the fast (inter-band) oscillations appear as well when more than one band is taken into account, but the nonlinear Bloch oscillations are less well represented. In particular the one-band cutoff in Coulomb gauge results only in a sinusoidal time-dependence in the current. As the number of bands is increased, both the dipolar and the Coulomb gauge description converge, but the results show that the dipolar gauge is advantageous over the Coulomb gauge for a few-band approximation in a gapped lattice system. 
 
The agreement of two-band dipolar approximation with the exact result becomes slightly worse at later times (Fig.~\ref{fig_curr}b). This is expected, as the electron is successively excited to higher bands in a process closely related to Landau Zener tunnelling. In order to investigate this excitation process in more detail, we consider the initial state $k=0,m=0$ subject to a constant electric field, $A(t)=-Et$, for a case with a smaller gap ($V_0=1$). Figure~\ref{fig_curr}h) shows the time-evolution of the probability for the electron to stay in the first band, obtained within the two-band approximation. In the dipolar gauge, states with different band index $m$ are mixed via the dipolar matrix element. Around time $t=\pi/E$, one has $k-qA(t)=-\pi$ (for $k=0$), so that the system passes through an anti-crossing where the energy difference $|\epsilon_{k-A(t),1}-\epsilon_{k-A(t),2}|$ is minimal. The reduction $\Delta P$ of the occupation probability in the first band during the traverse of the anticrossing  can be related to Landau-Zener tunnelling \cite{niu1998}, with $\Delta P=\exp[-\frac{\pi}{2}E_g^2/(2\pi E/a)]$ (black arrows in Fig.~\ref{fig_curr}h). We note that due to the gauge-dependence of momentum this probability does not correspond to the projection of the exact wave function to the equilibrium basis (eigenstates of Bloch momentum $k$), but fits well the exact result projected onto a basis which is defined by the eigenstates of gauge-invariant Bloch momentum $k_m=k-A(t)$. The two-band calculation in Coulomb gauge gives a completely different result for either of the two basis sets. In fact, in the Coulomb gauge the physical energy bands, defined with gauge-invariant Bloch momentum $k_m$, are time-dependent superpositions of several states in the bare momentum basis. During time-evolution, many orbitals are therefore needed to recover the gauge-invariant current and electron occupation in the lowest physical band. This clearly shows the advantage of preserving the gauge structure under the few-band approximation, which is satisfied by the dipolar Hamiltonian.

\subsection{Quantum case: Electron-polariton bands}

\subsubsection{Formulation}

As a second problem, we consider the one-dimensional chain coupled to the quantized modes of a perfectly isolated resonator. We assume that the resonator Hamiltonian $H_{em}$, Eq.~\eqref{lwcbas} has been diagonalized in the form $H_{em} = \sum_\nu \frac{\omega_\nu}{2} (Q_\nu^2 + \Pi_\nu^2)$, with canonical variables $Q_\nu$ and $\Pi_\nu$, and use the mode expansion described around Eq.~\eqref{nm,mnm,m}. In the following we take into account only one mode ($\nu\equiv0$). The mode function  $\bm b_0(\bm r)$ is assumed to be constant throughout the solid, and $b_0$ denotes the corresponding  component of the mode function along the chain direction.  Here the one mode approximation is intended to facilitate an exact comparison of the dipolar and Coulomb gauge. A quantitative solution of a given cavity setup may require more than one cavity mode, depending on the cavity geometry.

With one mode and the expansion \eqref{abkxlnm;z}, the Peierls factors $q\chi_{\alpha,\alpha'}$ in the dipolar Hamiltonian  \eqref{fghjfghjkfghj01} for sites at distance $|\bm R_\alpha - \bm R_\alpha'|=na$ become $ngQ_0$, with the dimensionless coupling constant
\begin{align}
\label{vbnm00000}
g= \frac{q b_0 a}{\hbar\sqrt{\omega_0 \epsilon_0}},
\end{align}
where $\hbar$ is restored for concreteness. With some manipulation, the dipolar light-matter Hamiltonian is then obtained from Eqs.~\eqref{fghjfghjkfghj01}- \eqref{fghjfghjkfghj03},
\begin{align}
\label{fhgjklkjhbvhbkl02}
H^{\rm Dip} &= \frac{\omega_0}{2}(Q_0^2+\Pi_0^2)+H_0 +  H_{EP} +H_{PP},
\end{align}
with
\begin{align}
H_0&=\sum_{k,m}
c_{k,m}^\dagger c_{k,m}
\epsilon_{k-gQ_0;m},
\\
H_{EP}
&=
\frac{g \omega_0  }{2}
\sum_{k,m,m'}
\Big[
c_{k,m}^\dagger c_{k,m'}
\Pi_0
D_{k-gQ_0;m,m'}
+
h.c.
\Big]
\\
H_{PP}
&=
\frac{g^2\omega_0  }{2}
\sum_{k,m,m'}
c_{k,m}^\dagger c_{k,m'}
\big(D_{k-gQ_0}^2\big)_{m,m'}.
\end{align}
($D_{k-gQ_0}^2$ implies a matrix multiplication in band indices.) Here only the first term of $H_{PP}$ [Eq.~\eqref{fghjfghjkfghj03}] has been included, because the second contribution is an electron-electron interaction  which vanishes for the case of one electron.

Similar to the semiclassical case, the projected Coulomb Hamiltonian is obtained as 
\begin{align}
\label{fhgjklkjhbvhbkl02}
&H^{\rm C} = \frac{\omega_0}{2}(Q_0^2+\Pi_0^2) +  
\sum_{k,m,m'}
c_{k,m}^\dagger c_{k,m'}
h^{\rm C}_{k;m,m'},
\\
&h^{\rm C}_{k;m,m'}=
\delta_{m,m'}\big[\epsilon_{k;m}
+\frac{g^2}{2} Q_0^2\big]
-
gQ_0
p_{k;m,m'}.
\end{align}
Below we compute the spectrum of the two Hamiltonians as a function of the band cutoff $m_{\rm max}$ and the coupling constant $g$. The wave function at momentum $k$ is expanded as 
\begin{align}
|\chi_k\rangle
=
\sum_{m=0}^{m_{\rm max}}
\sum_{n=0}^{n_{\rm max}}
\chi_{k;m,n} 
|k,m;n\rangle,
\end{align}
where $|k,m;n\rangle=\frac{1}{\sqrt{n!}}\big(a_0^\dagger\big)^n c_{k,m}^\dagger |0\rangle$ is the basis state with one electron in band $m$ and $n$ photons; $a_0^\dagger=\frac{1}{\sqrt{2}}(Q_0-i\Pi_0)$ is the photon creation operator. The numerical cutoff $n_{\rm max}$ in the number of photons is taken large enough  so that the spectrum is converged (up to $n_{\rm max}=100$, depending on the parameters). The dipolar Hamiltonian can become highly nonlinear in $Q_0$, and thus couple many photon states. Instead of a Taylor expansion in $Q_0$, we therefore calculate  matrix-elements directly in the $Q_0$ representation; for $X_k\equiv \epsilon_{k}, D_k$, 
\begin{align}
\label{ggegegeg}
\langle n|X_{k-gQ_0}|n'\rangle_{\rm ph}
=
\int dQ \,\psi^{\rm ph}_n(Q)\psi^{\rm ph}_{n'}(Q)X_{k-gQ},
\end{align}
where the eigenfunctions $\psi^{\rm ph}_n(Q) =e^{-Q^2/2}H_n(Q)/\sqrt{\pi}$ are given in terms of Hermite polynomials $H_n(Q)$.

\begin{figure}
 \includegraphics[width=0.9\columnwidth]{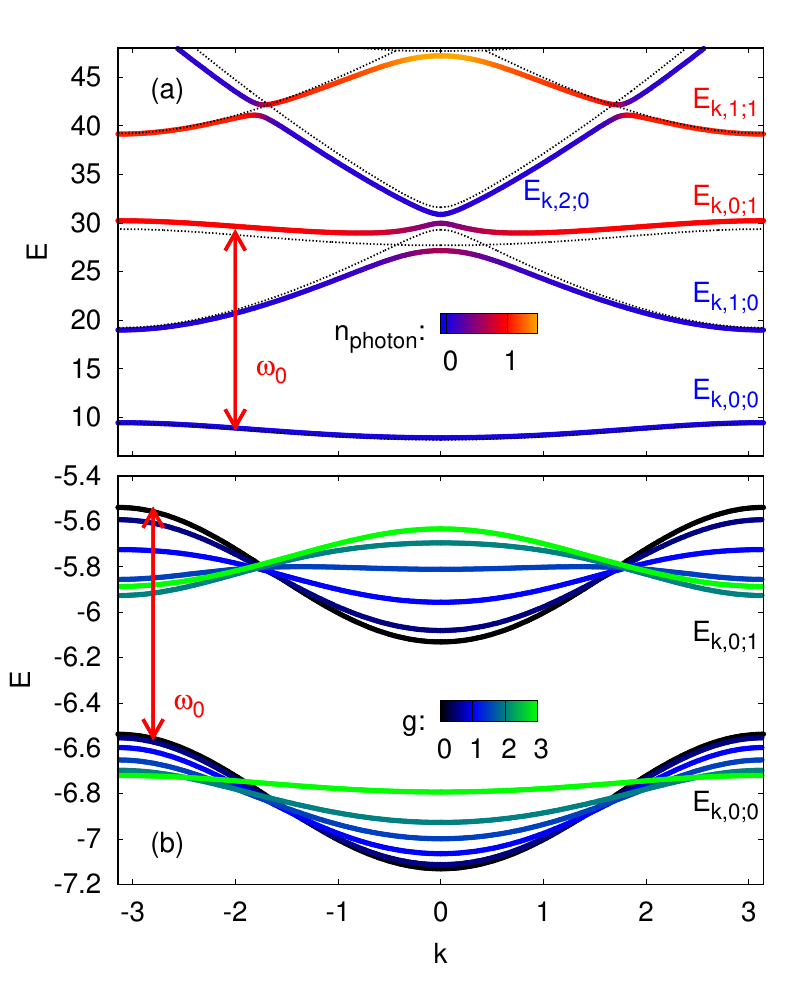}
\caption{
Electron-polariton band structure in different regimes.
 a) Photon energy resonant between bands, $V_0=5$, $\omega_0=20$.
Dashed lines correspond to $g=0$, colored lines to $g=1$. 
The color of the lines indicates the photon number expectation value $n_{\rm photon}=\langle \chi_{k;m;n}| a_0^\dagger a_0 | 
\chi_{k;m;n} \rangle$ in the states.
 b) Well-separated bands, photon energy off-resonant between bands, $V_0=10$, $\omega_0=1$.
 The lowest two electron-polariton bands are shown for different values of the coupling $g$.
}
\label{fig2}
\end{figure}

\subsubsection{Results}

We first analyze the exact band structure of the light-matter coupled system (i.e., the band and photon number cutoff is taken large enough such that the results are converged and identical in both gauges). Figure~\ref{fig2}a illustrates the case of well isolated bands and a photon energy which is resonant to the transition between the first and second band ($V_0=5$ and $\omega_0=20$). For $g=0$, the bands are given by the bare bands with  $n=0,1,2,...$ photons, $|k,m;n\rangle$, with energy $E_{k,m;n}^{(0)}=\epsilon_{k,0}+(n+\tfrac12)\omega_0$ (black dashed lines). For $g>0$, the wave function  $|k,m\;n\rangle$ adiabatically evolves into a hybridized electron-polariton band $|\chi_{k,m;n}\rangle$ with energy $E_{k,m;n}$. To illustrate the hybridization, the bands in Fig.~\ref{fig2}a are colored according to the photon number expectation value $\langle \chi_{k;m;n}| a_0^\dagger a_0 | \chi_{k;m;n} \rangle$. The hybridization opens a gap at the level crossings of the bands $E_{k,1;0}$ and $E_{k,0;1}$.

\begin{figure}
\includegraphics[width=0.9\columnwidth]{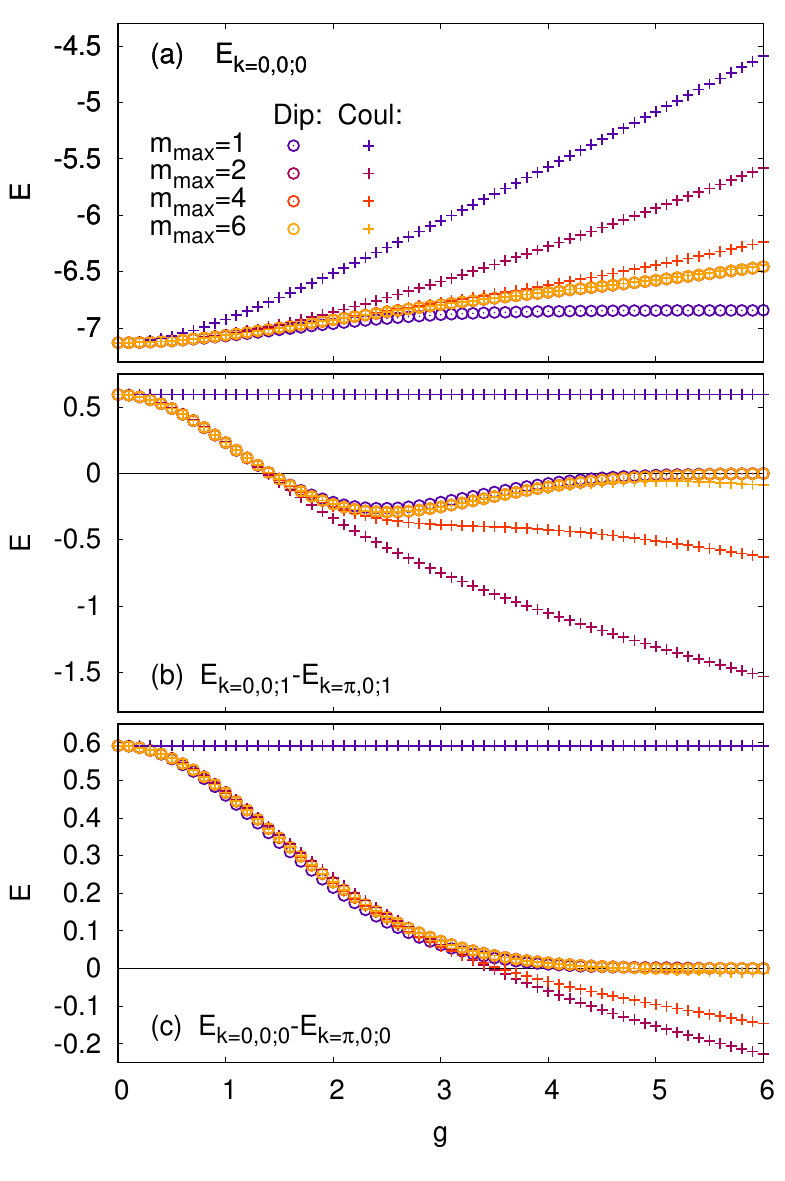}
\caption{
Convergence of the spectrum with the band cutoff $m_{\rm max}$ for the off-resonant case, for $V_0=10$, $\omega_0=1$.
a) Position of the lowest band. 
b) Width of the lowest (zero-photon) band $E_{k,0;0}$. 
c) Width of the one-photon band $E_{k,0;1}$. 
In all cases, the dipolar gauge is converged for $m_{\rm max}\ge 2$, so that symbols for $m_{\rm max}=2,4,6$ are indistinguishable.
}
\label{fig3}
\end{figure}

Figure~\ref{fig2}b shows a different parameter regime, with well isolated bands and off-resonant photon energy (same parameters $V_0=10$, $\omega_0=1$ as in Fig.~\ref{fig_curr}). Here we focus only on the dependence of the lowest two bands $E_{k,0;0}$ and $E_{k,0;1}$ on the coupling $g$. The first electron band $E_{k,1;0}$ is out of scale, c.f., Fig.~\ref{fig_curr}d). With increasing $g$, one observes both a shift of the bands and a renormalization of the dispersion in $k$. The shift of the bands is, to a large extent, given by the $g^2$ terms in the Hamiltonians, which derives from the $\bm A^2$ term in Coulomb gauge, and from the $H_{PP}$ term in dipolar gauge. This shows that it is  crucial to consistently take into account the  dipolar interactions $H_{PP}$ when switching between different gauges \cite{Keeling2007}. The band narrowing can be interpreted as the quantum analog of dynamical localization \cite{Dunlap1986}. This is most easily understood in dipolar gauge, where the narrowing is already accurately described in the one-band approximation (see below): In the classical case, when a system is driven with a time-periodic high-frequency field corresponding to $Q \to Q(t) = Q_{\rm max}\cos(\omega t)$, the narrowing arises from a time-average of the band structure over one period,
\begin{align}
\label{flippp}
\bar \epsilon_{k,0} = \frac{1}{T}\int_0^T dt\,  \epsilon_{k-gQ(t),0}. 
\end{align}
 In the limit of well-separated bands, where the band has the form $ \epsilon_{k,0}\approx -2J\cos(k)$, this leads to a renormalization $\bar \epsilon_{k,0} = \epsilon_{k,0} \mathcal{J}_0(gQ_{\rm max})$ of the band given by the zeroth order Bessel function. Equation \eqref{flippp} can be transformed to an average of $\epsilon_{k-gQ,0}$ over the classical probability of finding the oscillator at $Q$, 
\begin{align}
\label{flipppp}
\bar \epsilon_{k,0} = \frac{1}{\pi}\int dQ \frac{\theta(|Q_{\rm max}|-|Q|)}{\sqrt{Q_{\rm max}^2-Q^2}} \epsilon_{k-gQ,0}.
\end{align}
In the quantum case, the corresponding probability distribution is determined by the photon wave function. If additional inter-band dipolar terms are neglected, the lowest electron-polariton band becomes [c.f.~Eq.~\eqref{ggegegeg}]
\begin{align}
E_{k,0;0} \approx \int \frac{dQ}{\pi} e^{-Q^2}  \epsilon_{k-gQ,0} + \frac{\omega_0}{2}.
\end{align}
(The dipolar term leads to an admixture of higher photon number states to the electron-polariton.)

\begin{figure}
\includegraphics[width=1\columnwidth]{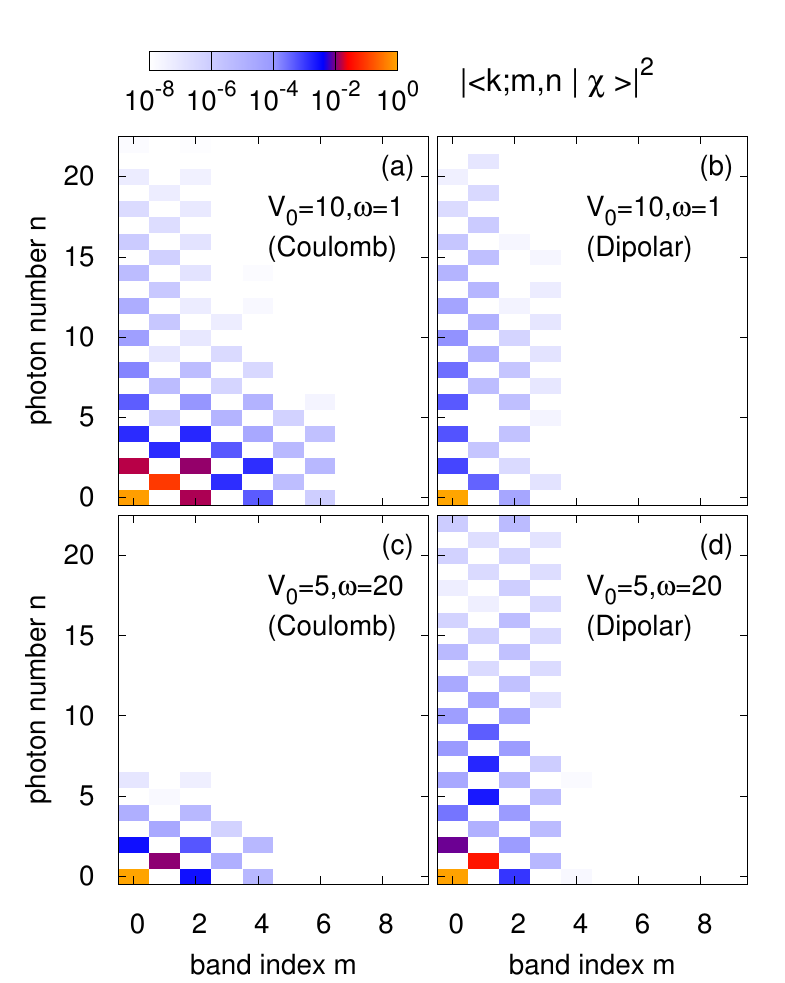}
\caption{
Electron-polariton wavefunction. The color shows the overlap $|\langle k,n;m|\chi\rangle|^2$ with the bare state $|k;m;n\rangle$ of band index $m$, photon number $n$.
a) and b) Off resonant case $V_0=10$, $\omega_0=1$.
c) and d) Resonant case $V_0=5$, $\omega_0=20$.
}
\label{fig4}
\end{figure}

In order to systematically analyze the convergence with the band cutoff $m_{\rm max}$, we focus on the renormalization of the lowest electron-polariton band in the large gap case ($V_0=10$, $\omega_0=1$). In Fig.~\ref{fig3}b and c we plot the splitting $E_{k=0,0;n}-E_{k=\pi,0;n}$ for the lowest two bands $n=0,1$ as a function of the coupling $g$, obtained in dipolar and Coulomb gauge with different band cutoff $m_{\rm max}$. The splitting can be taken as a measure of the band narrowing. For $n=1$, one even observes a sign change, analog to the band flipping for large arguments of the Bessel function in the classical case \eqref{flipppp}. For both $n=1$ and $n=0$, the band renormalization is essentially correct within the one-band picture in the dipolar gauge, while it is entirely missed in Coulomb gauge, where for $m_{\rm max}=1$ the bands are only shifted by the $g^2$ term. In Coulomb gauge, the band cutoff must be increased for larger coupling, consistent with the behavior in the semiclassical case, Fig.~\ref{fig_curr}. 

In Fig.~\ref{fig3}a we plot the position of the lowest band $E_{k=0,0;0}$. In the one-band approximation the shift is overestimated in Coulomb gauge and underestimated in dipolar gauge. This can be explained already to second order in $g$: In Coulomb gauge, the band shift arrises from the $g^2 Q_0^2$ contribution ($\bm A^2$ term), which is already present in the one-band approximation,  and is modified due to higher order contributions from the inter-band matrix elements $p_{k;m,m'}$ when more bands are taken into account. In the dipolar gauge, dipolar matrix elements vanish between states of the same band because of inversion symmetry, $D_{k;m,m}=0$. The band shifts arising due to $H_{PP}$  therefore become relevant only when more than one band is taken into account. 

The above result clearly show that the dipolar gauge is advantageous at least in the case of isolated bands. However, it should be noted that this statement includes only the convergence with the number of bands, not the number of photons. It is interesting to analyze the structure of the exact electron-polariton wave function in Coulomb gauge and dipolar gauge. Although the energy and the expectation value of gauge invariant observables is the same in both cases, the wave functions differ, as the two gauges correspond to a different light-matter basis. In Fig.~\ref{fig4} we plot the overlap $|\langle k,m;n|\chi_{k,0;0}\rangle|^2$ of the lowest electron-polariton state at $k=0$ and $g=2$, both in Coulomb gauge and in dipolar gauge. It is evident that the dipolar gauge is more localized in the band index, while the Coulomb gauge is more localized in the photon number. This can be expected because for large $g$ the dipolar gauge Hamiltonian becomes highly nonlinear in $Q_0$ and thus directly couples states with very different photon numbers. This observation is even more prominent at a level anti-crossing, see Fig.~\ref{fig4}c and d.

\section{conclusion}
\label{sec:555}

We have investigated few-band approximations to the light-matter Hamiltonian within different gauges, extending previous studies on the comparison between different gauges from atomic systems \cite{Bernardis2018, Di-Stefano2019} to the solid.  The general idea is to obtain tight-binding models from a projection of the continuum theory. When this projection  is performed in  different gauges, it effectively amounts to a projection to different light-matter hybrid states, because gauge transformations mix light and matter and therefore change the physical meaning of the bare electronic orbitals $c_\alpha$. This shows that the gauge choice can be crucial in order to derive the most accurate few band approximation.

 In particular, we have used a multi-center Power-Zienau-Woolley transformation to derive the dipolar gauge Hamiltonian, 
which features coupling to the vector potential via the Peierls phase factors and direct inter-band coupling through dipolar matrix elements coupling electronic states to the electric field. 
The dipolar Hamiltonian consistently exhibits advantages over the Coulomb-gauge Hamiltonian for the description of both the semi-classical dynamics and of quantum effects such as the band renormalization due to electron-polariton formation. Beyond that quantitative advantage, the semiclassical limit of the dipolar Hamiltonian has a particularly simple gauge structure  \cite{Boykin2001} as defined by Eq.~\eqref{fghjkl01}. The formalism also provides an alternative derivation applicable to the (already nontrivial) light-matter coupling in the semiclassical limit when more than one band, strong correlations, and strong laser fields beyond linear response have to be taken into account \cite{Golez2019}. 

In dipolar gauge, the relevant dipolar matrix elements may be determined ab-initio, chosen ad hoc to build some minimal model, or fitted to describe linear optical properties; in either case, the derivation of the dipolar-gauge Hamiltonian implies that the same dipolar matrix elements determine both the linear $\bm E\cdot \bm P$ light-matter interaction and the $\bm P^2$ interaction term when the modes of the cavity or the band is truncated. The importance of consistently keeping both terms  has recently been highlighted in the context of atomic physics \cite{Schaefer2019}. This consistency can also allow to perform further gauge transformations within the truncated model, changing, e.g., back to a Hamiltonian which contains only the vector potential but is nevertheless equivalent to the dipolar gauge. This procedure can be considered as implementing a non-linear truncation within the Coulomb gauge \cite{Di-Stefano2019}. 

The particular model which is used in the paper to compare dipolar and Coulomb gauge is, of course, rather simplistic. In particular, unless a certain resonance condition is at play, one would expect that all cavity modes, and not just one, contribute to the vacuum renormalization of the electron bands. On a fundamental level, it will be interesting to systematically take into account infinitely many cavity modes and thus perform the continuous limit from free space to the closed cavity. In the present manuscript, the restriction to one cavity mode rather serves to facilitate a comparison with the exact solution, while keeping physics qualitatively correct: The most dominant effect is a vacuum-induced band narrowing, which provides the quantum analog of dynamical localization in the classical case \cite{Dunlap1986}. That this effect is opposite to the cavity activated transport \cite{Schachenmayer2015} in a band which is non-dispersive at zero light-matter coupling.

At strong coupling, it is challenging to deal with the interacting problem of many electrons and photons, in any gauge. One can integrate out the photon field to obtain a description of the solid with induced retarded interactions, which can be dealt with (non-equilibrium) Green's function techniques. For this task, a gauge is favorable in which a linear coupling dominates. In Coulomb gauge, the coupling $\bm A \cdot \bm p$ is linear only in the weak coupling limit when the diamagnetic term is neglected. In the dipolar gauge, in contrast, the coupling is linear ($\bm E\cdot\bm P$) when the intra-band Peierls phase can be neglected. This may be true, e.g., when the bands are  are weakly dispersive in the direction of the field polarization. 

At strong coupling, even the basis set to which the projection is performed may itself be optimized. Together with the induced retarded interactions, this is a {\em downfolding } problem which is somewhat similar in nature to deriving few band models with a screened retarded interaction in correlated electron systems \cite{Biermann2014}. It will be interesting to see whether techniques developed in this context, 
can  be transferred to strongly coupled light-matter systems. 

\acknowledgments
We acknowledge discussions with D.~Jaksch, C. Sch\"afer, and Ph.~Werner. M.E. and J. Li were supported by the ERC starting grant No. 716648. GM and AG acknowledge the support of the European Research Council (ERC-319286-`QMAC'). 
GM acknowledges support from the FNS/SNF Ambizione Grant PZ00P2-186146.
AJM was supported by the Energy Frontier Research Center on Programmable Quantum Materials funded by the U.S. Department of Energy (DOE), Oce of Science, Basic Energy Sciences (BES), under award \# DE-SC0019443. The Flatiron Institute is a division of the Simons Foundation.

\appendix

\section{``Functional derivatives'' and Maxwell equations}

\label{AppendixA}

Below, we frequently need to calculate commutators $\big[O[\bm A],\bm \Pi(\bm r)\big]$ of an observable which depends only on $\bm A$, and has a support only inside the material of interest. An example is the matter part of the Hamiltonian \eqref{jjjjxsjsjsd}. We therefore consider a generic operator $\mathcal{O}[\bm A]$ which is a sum of terms
\begin{multline}
\mathcal{O}[\bm A]=
\int d^3\bm r_1 \cdots
d^3\bm r_n
\sum_{l_1,...,l_n}
K_{l_1,...,l_n}(\bm r_1,...,\bm r_n)\,\,\,\times
\\
\times \,\,\,
A_{l_1}(\bm r_1)\cdots A_{l_n}(\bm r_n),
\label{jdewk;wx;xw}
\end{multline}
and where $K_{l_1,...,l_n}(\bm r_1,...,\bm r_n)$ has a support only inside the matter, where the background dielectric has $\epsilon(\bm r)=1$. 
In the following, it will also be convenient to introduce a functional derivative defined by
\begin{multline}
\label{fefefllehjelkle}
\frac{\delta \mathcal{O}[\bm A]}{\delta A_l(\bm r)}=
\sum_{j=1}^n
\int d^3\bm r_1 \cdots
d^3\bm r_n
\sum_{l_1,...,l_n}
\delta(\bm r-\bm r_j)
\delta_{l_j,l}
\,\,\,\,\times\\
\times\,\,\,\,
K_{l_1,...,l_n}(\bm r_1,...,\bm r_n)
\prod_{a\neq j}
A_{l_a}(\bm r_a).
\end{multline}
When $\bm A$ is an unbounded operator, a true functional derivative with respect to $\bm A$ can be mathematically ill-defined. Equation~\eqref{fefefllehjelkle}, in contrast,  simply defines a shorthand notation for 
the well-defined canonical commutator as will be clear below.

Because the background dielectric is $\epsilon(\bm r)=1$ inside the matter,  field variables satisfy canonical communication relations in this region ($\hbar=1$),
\begin{align}
\label{,xsan;Z}
[A_j(\bm r'),\Pi_l(\bm r)]
=
i\delta^T_{jl}(\bm r-\bm r'),
\end{align}
with the transverse $\delta$-function $\delta^T_{jl}(\bm r-\bm r')$ which defines the projection on the transverse component of a field $\bm X=(X_1,X_2,X_3)$,
\begin{align}
\label{ehjjjekel;eee00}
X^T_l(\bm r) 
&= \sum_{j}\int d^3\bm r'\,\delta^T_{lj}(\bm r-\bm r') X_j(\bm r'),
\\
\delta^T_{lj}(\bm r)
&=
\frac{1}{V}\sum_{\bm q} e^{i\bm q\bm r}
\Big(
\delta_{lj}-\frac{q_lq_j}{\bm q^2}
\Big).
\end{align}
We therefore have the commutator
\begin{align}
[A_{l_1}(\bm r_1)
&
\cdots A_{l_n}(\bm r_n), \Pi_m(\bm r)]
\\
&=
\sum_{j=1}^n
[A_{l_j}(\bm r_j), \Pi_m(\bm r)]
\prod_{a\neq j}
A_{l_a}(\bm r_a)
\\
&=
i
\sum_{j=1}^n
\delta^T_{m,l_j}(\bm r-\bm r_j)
\prod_{a\neq j}
A_{l_a}(\bm r_a),
\end{align}
which can be denoted as 
\begin{align}
[\mathcal{O}[\bm A],\Pi_m(\bm r)]
=
i\int d^3\bm r'
\sum_{l}
\delta^T_{m,l}(\bm r-\bm r')
\frac{\delta \mathcal{O}[\bm A]}{\delta A_l(\bm r')}
\nonumber
\end{align}
with the shorthand notation \eqref{fefefllehjelkle}.
Using Eq.~\eqref{ehjjjekel;eee00}, we can finally denote
\begin{align}
\label{cwjlXA}
[\mathcal{O}[\bm A],\bm \Pi(\bm r)]
=
i\Big(\frac{\delta \mathcal{O}[\bm A]}{\delta \bm A(\bm r)}\Big)^T.
\end{align} 
Again we stress that this equation does not indicated the existence of a functional derivative with respect to an unbounded  operator $\bm A$, but merely defines a procedure to evaluate the commutator on the left-hand side  for observables that depend on $\bm A$ only. For example, for the derivative of the matter Hamiltonian \eqref{jjjjxsjsjsd}, this equation implies the usual relation
\begin{align}
\big[ H^C_{el}[\bm A],\bm \Pi(\bm r)\big]
=
i\Big(
\frac{\delta H^C_{el}}{\delta \bm A(\bm r)}
\Big)^T,
\end{align}
with 
\begin{align}
\frac{\delta H^C_{el}}{\delta \bm A(\bm r)}
&=
\frac{iq}{2m} \big[
(\bm \nabla\psi_{\bm r}^\dagger )  \psi_{\bm r}
-
\psi_{\bm r}^\dagger( \bm \nabla  \psi_{\bm r})
\big]
+\frac{q^2}{m} \bm A(\bm r)
\psi_{\bm r}^\dagger
\psi_{\bm r}
\nonumber\\
&\equiv
-\bm j_C(\bm r).
\end{align}

Finally, with Eq.~\eqref{cwjlXA}, one gets the resulting Heisenberg equations of motion for $\bm \Pi$ and $\bm \nabla\times\bm A$,
\begin{align}
\partial_t \bm \Pi(\bm r,t) &= i[H^{\rm C},\bm \Pi(\bm r,t)]
\nonumber
\\
&=
\label{ghjknbnm}
-\frac{1}{\mu_0}\bm\nabla\times\bm\nabla\times \bm A
-\Big(\frac{\delta H^{\rm C}_{el}[\bm A]}{\delta \bm A(\bm r)}\Big)^T,
\\
\partial_t \bm B(\bm r,t) 
&=
\partial_t \bm \nabla\times A(\bm r,t) 
=
\nabla\times 
i[H^{\rm C},\bm A(\bm r,t)]
\nonumber\\
&=
\bm \nabla\times \frac{1}{\epsilon_0\epsilon(\bm r)} \bm \Pi,
\end{align}
which correspond to the transverse Maxwell components of Maxwell equations with the identification \eqref{jhqvsv} and \eqref{kcwbslxNAZ}.

\section{The normal mode expansion}
\label{norm_exp}

Following the standard quantization procedure, the operator $\bm A$ and $\bm \Pi$ are expanded using normal modes 
$\bm\phi_\nu(\bm r)$ $Q_\nu,P_\nu$ in Eq.~\eqref{abkxlnm;z} and \eqref{hhhshwh}.
As in Ref.~\cite{Glauber1991}, we allow in general for complex expansion coefficients $Q_\nu,P_\nu$ .  %
 As the 
resulting $\bm{A}$ and $\bm \Pi$ have to be hermitian, the expansion must satisfy the following condition, 
\begin{align}
\sum_\nu\bm{\phi}_\nu(\bm{r})Q_\nu &= \sum_\nu\bm{\phi}^{*}_\nu(\bm{r})Q^\dag_\nu,\\
\sum_\nu\eta_\nu\bm{\phi}^*_\nu(\bm{r})\Pi_\nu &= \sum_\nu\eta_\nu\bm{\phi}_\nu(\bm{r})\Pi^\dag_\nu,
\end{align}
which allow $Q_\nu=Q^\dag_\nu$ and $\Pi_\nu=\Pi^\dag_\nu$ for real mode $\bm \phi_\nu=\bm \phi^*_\nu$, as the function 
$\bm{\phi}_\nu$'s are orthonormal. 
In general, the $Q_\nu,P_\nu$ must be non-hermitian, and should be related to creation and annihilation operators $a_\nu,a^\dag_\nu$ 
in the following way \cite{Glauber1991}
\begin{align}
Q_\nu&=\frac{1}{\sqrt{2}}\left(a_\nu+\sum_{\nu'}U^*_{\nu'\nu}a^\dag_{\nu'}\right),\\
\Pi_\nu&=i\frac{1}{\sqrt{2}}\left(a^\dag_\nu-\sum_{\nu'}U_{\nu\nu'}a_{\nu'}\right),
\end{align}
with the matrix $U_{\nu'\nu}=\eta_\nu\int\epsilon(\bm{r})\bm{\phi}_{\nu'}(\bm{r})\cdot\bm{\phi}_\nu(\bm{r})d^3\bm{r}$ which vanishes 
if frequency $\omega_\nu\ne\omega_{\nu'}$ since the eigenvalue problem~\eqref{nm,mnm,m} is invariant under 
complex conjugate, and $\bm{\phi}^*_\nu$ has the same frequency as $\bm{\phi}_\nu$. Using the creation and annihilation operators, the 
expansion \eqref{abkxlnm;z} and \eqref{hhhshwh} can be recast into the explicitly hermitian form, 
\begin{align}
        \bm{A}(\bm{r}) &= \frac{1}{\sqrt{2}}\sum_\nu [a_\nu \bm{\phi}_\nu(\bm{r}) + a^\dag_\nu\bm{\phi}^*_{\nu}(\bm{r})], \\
        \bm{\Pi}(\bm{r}) &= -\frac{i\epsilon(\bm{r})}{\sqrt{2}}\sum_\nu \eta_\nu[a_\nu \bm{\phi}_\nu(\bm{r}) - a^\dag_\nu\bm{\phi}^*_{\nu}(\bm{r})].
\end{align}
In principle, it is possible to expand the dipolar hamiltonian using this alternative representation of operators, leading to a slightly different way of 
mode summation in~\eqref{jbvbjbx01} and~\eqref{jbvbjbx02}. 
When cutting off normal modes in an approximate treatment, one has to consistently treat $\bm \phi_\nu$ and $\bm \phi^*_\nu$
 to maintain the realness of operators $\bm{A}$ and $\bm{\Pi}$. Fortunately, this condition is automatically satisfied for a truncation in the frequency 
 space, as $\bm{\phi}_\nu$ and $\bm{\phi}^*_\nu$ are degenerate in frequency
  (and thus $U_{\nu\nu'}\propto\delta(\omega_\nu-\omega_{\nu'})$). 

For the practical calculations in the main text, we have considered the simplest case where only one mode is retained and its amplitude $b_0$ 
is uniform in the solid. In that case, the $Q_0=(a_0+a^\dag_0)/\sqrt{2}$ and $P_0=i(a_0-a^\dag_0)/\sqrt{2}$ become identical to the dipole 
approximation. 

\section{Polarization operator Eq.~\eqref{jgkxalsz;} for a general gauge}
\label{sechdjd}

To derive the polarization operator Eq.~\eqref{jgkxalsz;}, we note that the unitary matrix can be written in the form
\begin{align}
\label{lqxn;maz}
W[\bm A]=e^{-iS[\bm A]},
\end{align}
where $ S$ is hermitian. We first note that the transformation Eq.~\eqref{lcnaxs} with \eqref{lqxn;maz} is generated by a $\mathcal{W}$ of the form
\begin{align}
\mathcal{W}=\exp\left(i\sum_{\alpha\alpha'}c^\dag_\alpha S[\bm A]_{\alpha\alpha'}c_{\alpha'}\right).
\end{align}
We can verify that $\mathcal{W}^\dag c_\alpha \mathcal{W}=\sum_{\alpha'} W[\bm A]_{\alpha\alpha'}c_{\alpha'}$ by using the 
Baker-Campbell-Hausdorff formula $e^B A e^{-B}=\sum_n\frac{1}{n!}[B,A]_n$ or by simply expanding $\mathcal{W}$. 

It is then straightforward to evaluate
\begin{align}
 \bar{\bm \Pi}&=\mathcal{W}\bm{\Pi}\mathcal{W}^\dag\nonumber\\
        &=\bm{\Pi} + i\sum_{\alpha\alpha'}c^\dag_{\alpha}\left[S_{\alpha\alpha'}[\bm{A}],\bm{\Pi}\right]c_{\alpha'}\nonumber\\
        &=\bm{\Pi} +\sum_{\alpha\alpha'}c^\dag_{\alpha}\left[\left(W^\dag[\bm{A}]\bm{\Pi}W[\bm{A}]\right)_{\alpha\alpha'}-\bm{\Pi}\delta_{\alpha\alpha'}\right] c_{\alpha'}\nonumber\\
        &=\bm{\Pi} +\sum_{\alpha\alpha'}c^\dag_{\alpha}\left(W^\dag\left[W,\bm{\Pi}\right] \right)_{\alpha\alpha'}c_{\alpha'},
         \nonumber\\
         &=\bm{\Pi}-i\sum_{\alpha\alpha'}\left(W^\dag\frac{\delta}{\delta \bm A(\bm r)}W\right)^T_{\alpha\alpha'}c^\dag_{\alpha}c_{\alpha'},
\end{align}
where the derivative with respect to $\bm A$ is an expression for the commutator as explained in Eq.~\eqref{cwjlXA}.
With the identification $\bm{M}[\bm A]=-iW^\dag\frac{\delta}{\delta \bm A(\bm r)}W$, this completes the proof of Eq.~\eqref{jgkxalsz;}. As usual, the 
functional derivative is evaluated using the corresponding commutator in all practical calculations.

\section{Polarization operator in the dipolar gauge}
\label{pppqqqppp01}

\subsubsection*{The matrix W, Eq.~\eqref{lcnaxs}, for the dipolar gauge }

In order to derive the Polarization $\bm P$ in the dipolar approximation, we first explicitly write down the matrix $W[\bm A]$ [Eq.~\eqref{lcnaxs}] corresponding to the transformation from Wannier orbitals $c_\alpha$ to the hybrid field matter orbitals $\bar c_\alpha$, and the use the general expression \eqref{jgkxalsz;} for the polarization.

The orbitals $\bar w_{\alpha}(\bm r)\approx \tilde w_{\alpha}(\bm r)$ [Eq.~\eqref{sbxalgg}] can be used to define  hybrid light-matter field operators,
\begin{align}
\label{weicqbx;'}
\bar c_\alpha
=
\int d^3\bm r\,
\bar w_{\alpha}(\bm r)^\dagger\,
\psi(\bm r),
\end{align}
which satisfy canonical anti-commutation relations 
\begin{align}
[\bar c_{\alpha}^\dagger,\bar c_{\alpha'}]_+=\delta_{\alpha,\alpha'}
\end{align}
because of the orthogonality. This implicitly defines a unitary transformation $\mathcal{W}$ such that
\begin{align}
\label{kaskb}
\bar c_\alpha
=
\mathcal{W}
c_\alpha
\mathcal{W}^\dagger.
\end{align}
With Eq.~\eqref{kaskb}, the field operator transforms as
\begin{align}
\bar \psi(\bm r)
&\equiv
\mathcal{W}
\psi(\bm r)
\mathcal{W}^\dagger
\stackrel{\eqref{xqQSKWKQK}}{=}
\sum_{\alpha} w_{\alpha}(\bm r) \bar c_{\alpha}
\nonumber\\
&
\stackrel{\eqref{weicqbx;'}}{=}
\sum_{\alpha}
\int d^3\bm r'\,
w_{\alpha}(\bm r)
\bar w_{\alpha}(\bm r')^\dagger\,
\psi(\bm r')
\nonumber\\
&
\stackrel{\eqref{xqQSKWKQK}}{=}
\sum_{\alpha,\alpha'}
\int d^3\bm r'\,
w_{\alpha}(\bm r)
\bar w_{\alpha}(\bm r')^\dagger\,
w_{\alpha'}(\bm r') c_{\alpha'}
\nonumber\\
&\approx
\sum_{\alpha,\alpha'}
\int d^3\bm r'\,
w_{\alpha}(\bm r)
e^{iq\chi(\bm r,\bm R_\alpha)}
w_{\alpha}(\bm r')^*\,
w_{\alpha'}(\bm r') c_{\alpha'}
\nonumber\\
&=
\sum_{\alpha}
e^{iq\chi(\bm r,\bm R_\alpha)}
w_{\alpha}(\bm r)
c_{\alpha}.
\label{jaxblnJSBDLna}
\end{align}
A projection of the first and last line on $w_\alpha^*$ gives 
\begin{align}
\label{gggsgsgsgs}
&\bar c_{\alpha}
=
\sum_{\alpha'}
W[\bm A]_{\alpha,\alpha'}c_{\alpha'}
\\
\label{scn;axz'mlxa}
&W[\bm A]_{\alpha,\alpha'}
=
\int d^3{\bm r}\,
w_{\alpha}(\bm r)^*
e^{iq\chi(\bm r,\bm R_{\alpha'})}
w_{\alpha'}(\bm r).
\end{align}
As $W$ is unitary in the matter indices (up to magnetic corrections), this is a transformation precisely of the form \eqref{lcnaxs}, so that all properties from the previous section App.~\ref{sechdjd} can be reused to determine the corresponding Hamiltonian $H_\mathcal{W}$, and the polarization operator $\bm P_\mathcal{W}$.  

\subsubsection*{The polarization density Eq.~\eqref{gegege}}

Using Eqs.~\eqref{jgkxalsz;01} and \eqref{scn;axz'mlxa} we have
\begin{align}
&\bm M(\bm r)_{\alpha,\alpha'}
=
-i
\sum_{\alpha''}
(W[\bm A]^\dagger)_{\alpha,\alpha''}
\frac{\delta}{\delta \bm A(\bm r)}
W[\bm A]_{\alpha'',\alpha'}
\nonumber\\
&=
-i
\sum_{\alpha''}
W[\bm A]_{\alpha'',\alpha}^\dagger
\frac{\delta}{\delta \bm A(\bm r)}
W[\bm A]_{\alpha'',\alpha'}
\nonumber\\
&\stackrel{\eqref{scn;axz'mlxa}}{=}
-i
\sum_{\alpha''}
\int d^3{\bm r'}\,
w_{\alpha''}(\bm r')
e^{-iq\chi(\bm r',\bm R_{\alpha})}
w_{\alpha}(\bm r')^*
\,\,\,\times\nonumber\\&\,\,\,\,\times\,\,\,\,
\frac{\delta}{\delta \bm A(\bm r)}
\int d^3{\bm r''}\,
w_{\alpha''}(\bm r'')^*
e^{iq\chi(\bm r'',\bm R_{\alpha'})}
w_{\alpha'}(\bm r'').
\end{align}
As the derivative acts only on $\chi(\bm r'',\bm R_{\alpha'})$, we can contract the $\sum_{\alpha''} w_{\alpha''}(\bm r')w_{\alpha''}(\bm r'')^*=\delta(\bm r''-\bm r')$, leading to
\begin{multline}
\bm M(\bm r)_{\alpha,\alpha'}
=
q\int d^3{\bm r'}\,
w_{\alpha}(\bm r')^*
e^{-iq\chi(\bm r',\bm R_{\alpha})}
\,\,\,\,\times\\\times\,\,\,\,
\frac{\delta \chi(\bm r',\bm R_{\alpha'})}{\delta \bm A(\bm r)}
e^{iq\chi(\bm r',\bm R_{\alpha'})}
w_{\alpha'}(\bm r').
\end{multline}
With the electric approximation $e^{-iq\chi(\bm r',\bm R_{\alpha})} e^{iq\chi(\bm r',\bm R_{\alpha'})} \approx e^{iq\chi_{\alpha,\alpha'}}$,
this gives 
\begin{align}
\label{ggsgggsg}
\bm M(\bm r)_{\alpha,\alpha'}
&=
q
e^{iq\chi_{\alpha,\alpha'}}
\int d^3{\bm r'}\,
w_{\alpha}(\bm r')^*
\frac{\delta \chi(\bm r',\bm R_{\alpha'})}{\delta \bm A(\bm r)}
w_{\alpha'}(\bm r').
\end{align}
To further simplify this expression, we use the expansion \eqref{abkxlnm;z} of $\bm A$ within the integral \eqref{jjwjwkqblq} for $\chi$, and perform the derivative \eqref{fghjfgfhhgh}. This leads to the expansion \eqref{kvxlnzA;}, where the expansion coefficients \eqref{fghjkljh} are given by  
\begin{align}
\label{maintext:ggsgggsg}
(\bm M_{\nu})_{\alpha,\alpha'}
&=
qe^{iq\chi_{\alpha,\alpha'}}
\int d^3{\bm r'}\,
w_{\alpha}(\bm r')^*
\chi^{\nu}_{\alpha'}(\bm r')
w_{\alpha'}(\bm r'),
\end{align}
with the straight path integral over the mode function,
\begin{align}
\chi^{\nu}_{\alpha'}(\bm r)
=
\int_{\bm R_{\alpha'}}^{\bm r}
d\bm r' \cdot \bm \phi_\nu(\bm r').
\end{align}
Further we assume that the relevant mode  functions vary little over one lattice spacing (this corresponds to the electric dipolar approximation), and thus replace $\bm \phi_\nu(\bm r')\approx \bm \phi_\nu(\bm R_{\alpha,\alpha'})$, with the position $\bm R_{\alpha,\alpha'}\equiv \frac{\bm R_{\alpha}+\bm R_{\alpha'}}{2}$ of the bond $(\alpha,\alpha')$. This gives 
\begin{align}
\label{App-gegege}
 P_\mathcal{W,\nu}
&=
q
\sum_{\alpha,\alpha'}
c_{\alpha}^\dagger c_{\alpha'}
e^{iq\chi_{\alpha,\alpha'}}
\big(
\bm \phi_\nu(\bm R_{\alpha,\alpha'})
\cdot
\bm D_{\alpha,\alpha'}\big),
\end{align}
within the expansion \eqref{kvxlnzA;},
with the dipolar Matrix elements
\begin{align}
\label{hbwhkqjs}
\bm D_{\alpha,\alpha'}
&=
\int d^3{\bm r}\,
w_{\alpha}(\bm r)^*
(\bm r-\bm R_{\alpha'})
w_{\alpha'}(\bm r).
\\
\label{cghjhbjknjkjn}
&=
\int d^3{\bm r}\,
w_{\alpha}(\bm r)^*
\,\bm r\,
w_{\alpha'}(\bm r).
\end{align} 
(The second line follows from the orthogonality of the Wannier orbitals.)
This is precisely Eq.~\eqref{gegege}. The expression is analogous to a coarse graining of the polarization density
\begin{multline}
\label{gegege01}
\bm P_\mathcal{W}(\bm r)
=
q
\sum_{\alpha,\alpha'}
c_{\alpha}^\dagger c_{\alpha'}
e^{iq\chi_{\alpha,\alpha'}}
\bm D_{\alpha,\alpha'}
\delta(\bm r- \bm R_{\alpha,\alpha'}).
\end{multline}
In general, one can carry out a multipolar expansion for \eqref{maintext:ggsgggsg} to include the higher order terms, such as the quadrupolar contribution. 

\section{Electronic Hamiltonian in dipolar gauge}
\label{pppqqqppp02}

Using the expansion \eqref{jaxblnJSBDLna} of the field operators $\bar \psi(\bm r)$ in $\mathcal{W}$ gauge, the kinetic energy of the continuum Hamiltonian \eqref{jjjjxsjsjsd} is written as
\begin{align}
\mathcal{W}
H_{el,C}^{(0)}
\mathcal{W}^\dagger
&=
\int d^3\bm r
\,\,
\bar \psi (\bm r)^\dagger
\frac{(-i\bm \nabla -q \bm A(\bm r))^2}{2m}
\bar \psi(\bm r)
\nonumber\\
&=
\sum_{\alpha,\alpha'}
c_{\alpha}^\dagger c_{\alpha'}
\bar h^{kin}_{\alpha,\alpha'},
\end{align}
with 
\begin{multline}
\bar h^{kin}_{\alpha,\alpha'}
=
c_{\alpha}^\dagger c_{\alpha'}
\int d^3\bm r
\,\,
w_{\alpha}(\bm r)^*
e^{-iq\chi(\bm r,\bm R_\alpha)}
e^{iq\chi(\bm r,\bm R_{\alpha'})}
\,\,\,\times\\\times\,\,\,\,
\frac{(-i\bm \nabla + q[\bm \nabla \chi(\bm r,\bm R_{\alpha'})]-q \bm A(\bm r))^2}{2m}
w_{\alpha'}(\bm r).
\end{multline}
From Eq.~\eqref{jjwjwkqblq} it follows that \cite{LoudonBook}
\begin{align}
\label{lxqnla}
\bm \nabla
\chi(\bm r,\bm 0)=
\bm A(\bm r) 
-
\int d^3\bm r'\,\bm \theta(\bm r,\bm r')\times\bm B(\bm r'),
\end{align}
with 
\begin{align}
\bm \theta(\bm r,\bm r')
=
-\int_0^1 ds \, s\, \bm r' \delta(\bm r - s\bm r').
\end{align}
Hence the PZW transformation removes the vector potential from the kinetic energy, up to small magnetic terms of the order of the flux per lattice plaquette. Neglecting again these magnetic contributions, and consistently using the electric approximation $e^{-iq\chi(\bm r,\bm R_\alpha)}e^{iq\chi(\bm r,\bm R_{\alpha'})}\approx e^{iq\chi_{\alpha\alpha'}}$ for the phase factors, we have
\begin{align}
\bar h^{kin}_{\alpha,\alpha'}
=
e^{iq\chi_{\alpha,\alpha'}}
\int d^3\bm r
\,
w_{\alpha}(\bm r)^*
\frac{(-i\bm \nabla )^2}{2m}
w_{\alpha'}(\bm r),
\end{align}
i.e., the matrix element is given by the field-free matrix element dressed by a Peierls phase. 

To show that the analogous Peierls substitution can be made for all other matrix elements, it suffices to consider the transformation of an operator which is a function of position, $O=\int d^3\bm r\,\,\psi (\bm r)^\dagger O(\bm r)\psi(\bm r)$. This covers both matrix elements of the lattice potential and of the Coulomb interaction, the latter involving a pair of such operators. Using again the expansion \eqref{jaxblnJSBDLna}, the transformed operator is written as $\mathcal{W} O \mathcal{W}^\dagger = \sum_{\alpha,\alpha'}
c_{\alpha}^\dagger c_{\alpha'} \bar o_{\alpha,\alpha'}$, with 
\begin{align}
\bar o_{\alpha,\alpha'}
&=
\int d^3\bm r
\,\,
w_{\alpha}(\bm r)^*
e^{-iq\chi(\bm r,\bm R_\alpha)}
O(\bm r)
e^{iq\chi(\bm r,\bm R_{\alpha'})}
w_{\alpha'}(\bm r).
\end{align}
With the electric approximation, $e^{-iq\chi(\bm r,\bm R_\alpha)}e^{iq\chi(\bm r,\bm R_{\alpha'})}\approx e^{iq\chi_{\alpha\alpha'}}$, this expression becomes the Peierls substitution $\bar o_{\alpha,\alpha'}=e^{iq\chi_{\alpha,\alpha'}}o_{\alpha,\alpha'}$ of the field-free matrix elements
\begin{align}
o_{\alpha,\alpha'}
&=
\int d^3\bm r
\,\,
w_{\alpha}(\bm r)^*
O(\bm r)
w_{\alpha'}(\bm r).
\end{align}

\section{Details of the one-dimensional solid}
\label{app:fghjkl}

In this section we provide some details for the solution of the model system defined by Eq.~\eqref{modelAAA}.
Eigenstates of the un-driven Hamiltonian 
\begin{align}
H=\frac{p_x^2}{2}+2V_0\cos\left(Gx\right).
\end{align}
are Bloch bands $\langle x|k,m\rangle=\phi_{k,m}(x)$ with quasi-momentum $k\in[-\pi,\pi)$ and band energy $\epsilon_{k,m}$, and band index $m=0,1,2,3,...$ To determine these functions, we use a plane wave representation of the Bloch states, $\phi_{k,m}(x)=\frac{1}{\sqrt{L_x}}\sum_{n}u_{k,m}(n)e^{i(k+nG)x}$, leading to the eigenvalue problem
\begin{multline}
\epsilon_{k,m} u_{k,m}(n)=\frac{(k+nG)^2}{2}u_{k,m}(n)
\\+V_0[u_{k,m}(n+1)+u_{k,m}(n-1)].
\end{multline}
In the numerical calculation the summation in the plane-wave expansion restricted to  $n\in \left[-\frac{n_{\rm max}}{2},\frac{n_{\rm max}}{2} \right]$  with a cutoff $n_{\rm max}=50$.  
Similarly, the exact time-dependent calculation can be carried out straightforwardly on the plane-wave basis, with the ansatz $\psi_{k}(x,t)=\frac{1}{\sqrt{L_x}}\sum_{n}u_{k}(n,t)e^{i(k+nG)x}$, with
\begin{multline}
i\dot u_{k}(n)=\frac{(k+nG-qA(t))^2}{2}u_{k}(n)
\\
+V_0[u_{k}(n+1)+u_{k}(n-1)].
\end{multline}

Dipolar matrix elements \eqref{cghjhbjknjkjn} in the Bloch basis become diagonal in momentum $k$ and can be determined as
\begin{align}
D_{k;m,m'}=i\sum_{n}u^*_{k,m}(n)\partial_k u_{k,m'}(n),
\end{align}
which is is well-defined provided the phases of complex $u_{k,m}$'s are fixed.
The bare momentum operator is given by
\begin{align}
p_{k;m,m'}=k\delta_{mm'}+\sum_{n} u^*_{k,m}(n) nG u_{k,m'}(n).
\end{align}

\bibliographystyle{apsrev4-1}
\bibliography{apssamp01}

\end{document}